\documentclass[12pt]{article}
\usepackage{amsmath}
\usepackage{amssymb}
\usepackage{amsfonts}
\usepackage{hyperref}
\usepackage[numbers,sort&compress]{natbib}

\allowdisplaybreaks
\begin{document}

\author{Constantin Bizdadea\thanks{%
E-mail: bizdadea@central.ucv.ro}, Solange-Odile Saliu\thanks{%
E-mail: osaliu@central.ucv.ro} \\
Department of Physics, University of Craiova\\
13 Al. I. Cuza Str., Craiova 200585, Romania}
\title{A novel mass generation scheme for an Abelian vector field}
\date{}
\maketitle

\begin{abstract}
A novel mass generation procedure for an Abelian vector field is proposed.
This procedure is based on the construction of a class of gauge theories
whose free limit describes a free massless vector field and a set of
massless real scalar fields by means of the antifield-BRST deformation
technique. The relationship between our results and those arising from the
Higgs mechanism based on the spontaneous symmetry breaking of an Abelian
gauge symmetry is emphasized. Some examples with one, two, and three scalars
are given.

\end{abstract}

\section{Introduction\label{intro}}

It is commonly believed that the only possible way to generate vector field
masses is the Higgs mechanism based on spontaneous symmetry breaking~\cite%
{Englert,Higgs1,Higgs2,Kibble}. The starting point of the simplest Abelian
version of the Higgs mechanism is the Lagrangian action (expressed in terms
of two real scalars)
\begin{align}
W_{0}[A^{\mu },\varphi _{1},\varphi _{2}]=\int d^{4}x\Big[& -\tfrac{1}{4}%
F_{\mu \nu }F^{\mu \nu }+\tfrac{1}{2}(D_{\mu }\varphi _{1})D^{\mu }\varphi
_{1}  \notag \\
& +\tfrac{1}{2}(D_{\mu }\varphi _{2})D^{\mu }\varphi _{2}-V(\varphi
_{1},\varphi _{2})\Big],  \label{1H}
\end{align}
where
\begin{align}
F_{\mu \nu }& =\partial _{\mu }A_{\nu }-\partial _{\nu }A_{\mu },  \label{2H}
\\
D_{\mu }\varphi _{1}& =\partial _{\mu }\varphi _{1}-q\varphi _{2}A_{\mu
},\qquad D_{\mu }\varphi _{2}=\partial _{\mu }\varphi _{2}+q\varphi
_{1}A_{\mu },  \label{2Hx} \\
V(\varphi _{1},\varphi _{2})& =\tfrac{1}{2}\mu ^{2}\big(\varphi
_{1}^{2}+\varphi _{2}^{2}\big)+\tfrac{1}{16}\lambda \big(\varphi _{1}^{2}
+\varphi _{2}^{2}\big)^{2}  \label{3H}
\end{align}
and the real constants $\mu ^{2}$ and $\lambda $ are taken such that $\mu
^{2}<0$ and $\lambda >0$. Action (\ref{1H}) is invariant under the gauge
transformations
\begin{equation}
\delta _{\epsilon }A^{\mu }=\partial ^{\mu }\epsilon ,\qquad \delta
_{\epsilon }\varphi _{1}=q\varphi _{2}\epsilon ,\qquad \delta _{\epsilon }
\varphi _{2}=-q\varphi _{1}\epsilon .  \label{3Hx}
\end{equation}
In this setting we find that the potential of the form (\ref{3H}) has an
absolute minimum for $\sqrt{\varphi _{1}^{2}+\varphi _{2}^{2}}=\sqrt{-4\mu
^{2}/\lambda }\equiv v_{0}$. Introducing some new fields defined by $\tilde{%
\varphi}_{1}=\varphi _{1}-v_{0}$ and $\tilde{\varphi}_{2}=\varphi _{2}$
(whose associated field operators display zero vacuum expectation values)
and reformulating relations (\ref{1H}) and (\ref{3Hx}) accordingly, we find
the action
\begin{align}
W_{0}[A^{\mu },\tilde{\varphi}_{1},\tilde{\varphi}_{2}]=\int d^{4}x\Big[& -%
\tfrac{1}{4}F_{\mu \nu }F^{\mu \nu }+\tfrac{1}{2}q^{2}v_{0}^{2}A_{\mu }
A^{\mu }  \notag \\
& +\tfrac{1}{2}(\partial _{\mu }\tilde{\varphi}_{1})\partial ^{\mu }\tilde{%
\varphi}_{1}+\mu ^{2}\tilde{\varphi}_{1}^{2}+\tfrac{1}{2}(\partial _{\mu }
\tilde{\varphi}_{2})\partial ^{\mu }\tilde{\varphi}_{2}  \notag \\
& -\tfrac{1}{16}\lambda \big(\tilde{\varphi}_{1}^{2}+\tilde{\varphi}_{2}^{2} %
\big)\big(\tilde{\varphi}_{1}^{2}+\tilde{\varphi}_{2}^{2} +4v_{0}\tilde{%
\varphi}_{1}\big)  \notag \\
& +qA_{\mu }\big(\tilde{\varphi}_{1}\partial ^{\mu }\tilde{\varphi}_{2} -%
\tilde{\varphi}_{2}\partial ^{\mu }\tilde{\varphi}_{1}\big)+qv_{0}A_{\mu }
\partial ^{\mu }\tilde{\varphi}_{2}  \notag \\
& +\tfrac{1}{2}q^{2}\big(\tilde{\varphi}_{1}^{2}+\tilde{\varphi}
_{2}^{2}+2v_{0}\tilde{\varphi}_{1}\big)A_{\mu }A^{\mu }\Big],  \label{4H}
\end{align}
invariant under the gauge transformations
\begin{equation}
\delta _{\epsilon }A^{\mu }=\partial ^{\mu }\epsilon ,\qquad \delta
_{\epsilon }\tilde{\varphi}_{1}=q\tilde{\varphi}_{2}\epsilon ,\qquad \delta
_{\epsilon }\tilde{\varphi}_{2}=-q(\tilde{\varphi}_{1}+v_{0})\epsilon .
\label{5H}
\end{equation}
Formula (\ref{4H}) may be synthesized into: (a) the vector field $A^{\mu }$
acquires the mass $\sqrt{q^{2}v_{0}^{2}}$; (b) the scalar field $\tilde{%
\varphi}_{1}$ (the Higgs boson) becomes massive, with the mass equal to $%
\sqrt{-2\mu ^{2}}$; (c) the scalar field $\tilde{\varphi}_{2}$ (the
Goldstone boson) is massless. The previous conclusions are involved by the
existence of the solution $v_{0}$ that minimizes the potential (\ref{3H}).

Let us consider now the case where $\mu ^{2}$ is arbitrary and $\lambda >0$
in formula (\ref{1H}). If we add to action (\ref{1H}) the functional
\begin{align}
\tilde{W}_{0}[A^{\mu },\varphi _{1},\varphi _{2}]= \int d^{4}x \Big[ &
qvA_{\mu }\partial ^{\mu }\varphi _{2}+\tfrac{1}{2} q^{2} v^{2} A_{\mu }
A^{\mu } +q^{2}v\varphi _{1}A_{\mu }A^{\mu }  \notag \\
& -\tfrac{1}{4}\lambda v\varphi _{1} \big( \varphi _{1}^{2}+\varphi _{2}^{2} %
\big) -\tfrac{1}{8} \lambda v^{2} \big( 3\varphi _{1}^{2} +\varphi _{2}^{2} %
\big)  \notag \\
& -v \Big( \mu ^{2}+\tfrac{1}{4}\lambda v^{2}\Big)\varphi _{1}\Big],
\label{6H}
\end{align}
where $v$ is an arbitrary, nonvanishing real constant, then we find that
action
\begin{equation}
\bar{W}_{0}[A^{\mu },\varphi _{1},\varphi _{2}]=W_{0}[A^{\mu },\varphi
_{1},\varphi _{2}]+\tilde{W}_{0}[A^{\mu },\varphi _{1},\varphi _{2}],
\label{7H}
\end{equation}
is invariant under the gauge transformations
\begin{equation}
\delta _{\epsilon }A^{\mu }=\partial ^{\mu }\epsilon , \qquad \delta
_{\epsilon }\varphi _{1}=q\varphi _{2}\epsilon , \qquad \delta _{\epsilon }
\varphi _{2}=-q(\varphi _{1}+v)\epsilon .  \label{8H}
\end{equation}
We mention that in (\ref{7H}) and (\ref{8H}) there is no a priori relation
among the constants $\mu ^{2}$, $\lambda $, and $v$. Thus, expression (\ref%
{7H}) emphasizes that the mass of the vector field $A^{\mu }$, $\sqrt{%
q^{2}v^{2}}$, is independent both of $\mu ^{2}$ and $\lambda $. In fact, the
constants $\mu ^{2}$ and $\lambda $ are involved in (\ref{7H}) in the
self-interactions and (possibly) some mass terms of the scalar fields. Let
us take a fixed value of $v$, say $\bar{v}$. If $\mu ^{2}\geq 0$, then the
scalars $\varphi _{1}$ and $\varphi _{2}$ are massive, their masses being $%
\sqrt{\mu ^{2}+(3/4)\lambda \bar{v}^{2}}$ and $\sqrt{\mu ^{2}+(1/4)\lambda
\bar{v}^{2}}$, respectively. Let us analyze now the case $\mu ^{2}<0$. If $%
\mu ^{2}$ and $\lambda $ satisfy the inequality $\big(-4\mu ^{2}/\lambda %
\big)<\bar{v}^{2}$, then the two scalars remain massive and their masses are
precisely those from the previous situation. If $\big(-4\mu ^{2}/\lambda %
\big)>\bar{v}^{2}>\big(-4\mu ^{2}/3\lambda \big)$, then the scalar $\varphi
_{1}$ remains massive (the value of its mass being the same from the above
situations) whereas the quantity $(-1/2)[\mu ^{2}+(1/4)\lambda \bar{v}^{2}]$
(that multiplies $\varphi _{2}^{2}$) should be regarded as a parameter, and
so on. Moreover, it is simple to see that if $\mu ^{2} +(1/4) \lambda \bar{v}%
^{2}=0$ ($\Leftrightarrow \bar{v}=v_{0}$), then the gauge theory described
by (\ref{7H}) and (\ref{8H}) reduces to that governed by relations (\ref{4H}%
) and (\ref{5H}) modulo the identifications $\varphi _{1}\leftrightarrow
\tilde{\varphi}_{1}$, $\varphi _{2} \leftrightarrow \tilde{\varphi}_{2}$.
These considerations argue that relations (\ref{7H}) and (\ref{8H}) underlie
a more general class of gauge theories than that corresponding to formulas (%
\ref{4H}) and (\ref{5H}).

The previous discussion raises the following problem: is there a procedure,
different from the Higgs mechanism, by which one may generate mass for a
vector field in the context of its interactions to an arbitrary set of real
scalar fields? The aim of this paper is to investigate the above problem. In
view of this, we implement the following steps: (i) we start from a free
theory in $D=4$ whose Lagrangian action is expressed like the sum between
the Maxwell action for a single vector field and that for a (finite)
collection of massless real scalar fields; (ii) we construct a general class
of gauge theories whose free limit is that from step (i) by means of the
deformation of the solution to the master equation~\cite{PLB1993,CM1998}
with the help of local BRST cohomology~\cite{CMP1995a,CMP1995b,PR2000}. On
the one hand, the procedure described so far does not account in any way for
the Higgs mechanism. On the other hand, it will be proved to produce the
next results: (iii) the vector field acquires mass irrespective of the
number of scalar fields from the collection; (iv) the gauge transformations
are deformed with respect to those from the free limit, but the associated
gauge algebra remains Abelian; (v) the propagator of the massive vector
field emerging from the gauge-fixed action behaves, in the limit of large
Euclidean momenta, like that from the massless case. In this way, the answer
to the investigated problem is affirmative. In the meantime, the method
based on steps (i) and (ii) enables a proper comparison with the Higgs
mechanism. In this context we show that our approach: (vi) is a
cohomological extension of the Abelian Higgs mechanism; (vii) reveals an
appropriate interpretation of the Higgs mechanism in the framework of the
BRST symmetry. Outcomes (iii)--(vii) stand for the main results of our paper.

We stress that, although the antifield-BRST deformation method is well known~%
\cite{PLB1993,CM1998}, its application to a free theory with an Abelian
vector field and a set of massless real scalar fields with the aim of
generating mass for the vector field in mind has not been approached so far.
This represents the core novelty of our scheme.

The paper is organized into nine sections. In Section~\ref{model} we
construct the antifield-BRST symmetry of the free theory. Section \ref%
{deform} briefly reviews the antifield-BRST deformation procedure. In
Section~\ref{master} we compute the deformed solution to the master equation
for the theory under consideration in the presence of some standard
hypotheses from field theory. The identification of the class of interacting
gauge theories is developed in Section~\ref{inter}. In Section~\ref{Higgs}
we focus on the comparison between the Abelian Higgs mechanism and our
procedure, while in Section~\ref{interpret} we give aninterpretation of the
Abelian Higgs mechanism in the light of the antifield-BRST symmetry. Section~%
\ref{examples} is devoted to the exemplification of our general results to
three particular cases. Section~\ref{concl} closes the paper with the main
conclusions.

\section{BRST symmetry of the free theory\label{model}}

We start with a Lagrangian action written as the sum between the action of
an Abelian vector field $A^{\mu }$ and that describing a finite set of
massless real scalar fields $\{ \varphi ^{A} \} _{A=\overline{1,N_{0}}}$
\begin{align}
S_{0}^{\mathrm{L}} \big[ A^{\mu },\varphi ^{A} \big]& =\int d^{4}x \Big[ -%
\tfrac{1}{4} F_{\mu \nu } F^{\mu \nu } +\tfrac{1}{2} k_{AB} \big( \partial
_{\mu } \varphi ^{A} \big) \partial ^{\mu }\varphi ^{B} \Big]  \notag \\
& \equiv S_{0}^{\mathrm{L,Maxwell}}[A_{\mu }] +S_{0}^{\mathrm{L,scalar}} %
\big[ \varphi ^{A} \big],  \label{2}
\end{align}
where the Abelian field strength is like in (\ref{2H}). We work with a
mostly negative metric in a Minkowski spacetime of dimension $D=4$, $\sigma
^{\mu \nu }=\sigma _{\mu \nu }=(+---)$ and a metric tensor $k_{AB}$ with
respect to the matter field indices (constant, symmetric, invertible, and
positively defined), $\varphi _{A}=k_{AB}\varphi ^{B}$. In this context, the
elements of its inverse will be symbolized by $k^{AB}$. It is easy too see
that the number of physical degrees of freedom of the starting theory is
equal to $N_{0}+2$.

Action (\ref{2}) is invariant under the gauge transformations
\begin{equation}
\delta _{\epsilon }A^{\mu }=\partial ^{\mu }\epsilon ,\qquad \delta
_{\epsilon }\varphi ^{A}=0,\quad A=\overline{1,N_{0}},  \label{2h}
\end{equation}%
that are Abelian and irreducible (independent). The previous properties
combined with the linearity of the field equations following from action (%
\ref{2}) in all fields allow us to conclude that the overall free model
under consideration is a linear gauge theory with a definite Cauchy order,
equal to two.

The construction of the antifield-BRST symmetry~\cite%
{BVPLB81,BVPLB83,BVPRD83,BVNPB84,CMP1990a,NPPB1990,Princeton1992,PR1995,IJGMMP1995,NPPB1997}
for this free theory starts with the identification of the algebra on which
the BRST differential $s$ acts. The generators of the BRST algebra are of
two kinds: fields/ghosts and antifields. The ghost spectrum for the model
under study reduces to the fermionic ghost $\eta $ associated with the gauge
parameter $\epsilon $ from (\ref{2h}). The antifield spectrum is organized
into the antifields $\{ A_{\mu }^{\ast },\varphi _{A} ^{\ast } \} $ of the
original fields together with the antifield of the ghost, $\eta ^{\ast }$.
The Grassmann parity ($\varepsilon $) of the BRST generators reads
\begin{align}
\varepsilon (A^{\mu }) &= \varepsilon \big( \varphi ^{A} \big) =0, &
\varepsilon (\eta ) &=1,  \label{G1} \\
\varepsilon (A_{\mu }^{\ast }) &= \varepsilon (\varphi _{A}^{\ast }) =1, &
\varepsilon (\eta ^{\ast }) &=0.  \label{G2}
\end{align}
Since the gauge generators from (\ref{2h}) are field-independent, it follows
that the BRST differential $s$ simply reduces to
\begin{equation}
s=\delta +\gamma ,  \label{3h}
\end{equation}
where $\delta $ signifies the Koszul--Tate differential, graded by the
antifield number $\mathrm{agh} $ ($\mathrm{agh} (\delta ) =-1$) and $\gamma $
stands for the longitudinal exterior derivative (in this case a true
differential), whose degree is named pure ghost number $\mathrm{pgh} $ ($%
\mathrm{pgh} (\gamma ) =1$). These two degrees do not interfere ($\mathrm{agh%
} (\gamma ) =0$, $\mathrm{pgh} (\delta )=0$). The overall degree that grades
the BRST algebra is known as the ghost number ($\mathrm{gh} $) and is
defined like the difference between the pure ghost number and the antifield
number, such that $\mathrm{gh} (s) =\mathrm{gh} (\delta ) =\mathrm{gh}
(\gamma ) =1$. According to the standard rules of the BRST method, the
corresponding degrees of the generators from the BRST algebra are valued
like
\begin{align}
\mathrm{agh} (A^{\mu }) &=0, & \mathrm{agh} \big( \varphi ^{A} \big) &=0, &
\mathrm{agh} (\eta ) &=0,  \label{4h} \\
\mathrm{agh} (A_{\mu }^{\ast }) &=1, & \mathrm{agh} (\varphi _{A}^{\ast })
&=1, & \mathrm{agh} (\eta ^{\ast }) &=2,  \label{5h} \\
\mathrm{pgh} (A^{\mu }) &=0, & \mathrm{pgh} (\varphi ^{A}) &=0, & \mathrm{pgh%
} (\eta ) &=1,  \label{6h} \\
\mathrm{pgh} (A_{\mu }^{\ast }) &=0, & \mathrm{pgh} (\varphi _{A}^{\ast })
&=0, & \mathrm{pgh} (\eta ^{\ast }) &=0.  \label{7h}
\end{align}
The actions of $\delta $ and $\gamma $ on the BRST generators that enforce
the fundamental cohomological requirements of the antifield BRST theory~\cite%
{CMP1990a,NPPB1990,Princeton1992,PR1995,IJGMMP1995,NPPB1997} are given by
\begin{align}
\delta A^{\mu } &=0, & \delta \varphi ^{A} &=0, & \delta \eta &=0,
\label{8h} \\
\delta A_{\mu }^{\ast } &= \partial ^{\nu }F_{\mu \nu }, & \delta \varphi
_{A}^{\ast } &= k_{AB}\partial _{\mu }\partial ^{\mu }\varphi ^{B}, & \delta
\eta ^{\ast } &= -\partial ^{\mu }A_{\mu }^{\ast },  \label{9h} \\
\gamma A^{\mu } &= \partial ^{\mu }\eta , & \gamma \varphi ^{A} &=0, &
\gamma \eta &=0,  \label{10h} \\
\gamma A_{\mu }^{\ast } &=0, & \gamma \varphi _{A}^{\ast } &=0, & \gamma
\eta ^{\ast } &=0,  \label{11h}
\end{align}
where both operators were taken to act like right derivations.

The Lagrangian BRST differential admits a canonical action in a structure
named antibracket and defined by decreeing the fields/ghosts conjugated with
the corresponding antifields, $s\cdot =(\cdot ,S)$, where $(,)$ signifies
the antibracket and $S$ denotes the canonical generator of the BRST
symmetry. It is a bosonic functional of ghost number zero, involving both
field/ghost and antifield spectra, that obeys the master equation
\begin{equation}
(S,S)=0.  \label{12h}
\end{equation}
The master equation is equivalent to the second-order nilpotency of $s$ and
its solution, $S$, encodes the entire gauge structure of the associated
theory. The solution to the master equation for the free model under study
takes the simple form
\begin{equation}
S=S_{0}^{\mathrm{L}}\big[A^{\mu },\varphi ^{A}\big]+\int d^{4}x\,A_{\mu
}^{\ast }\partial ^{\mu }\eta .  \label{13h}
\end{equation}

\section{Deformation procedure: a brief review\label{deform}}

Now, we consider the problem of consistent interactions that can be
introduced in gauge field theories in such a way that the couplings preserve
the original number of independent gauge symmetries. This matter is
addressed by means of reformulating the problem of constructing consistent
interactions as a deformation problem of the solution to the master equation
corresponding to a given \textquotedblleft free\textquotedblright\ theory~%
\cite{PLB1993,CM1998} in the framework of the local BRST cohomology~\cite%
{CMP1995a,CMP1995b,PR2000}. Such a reformulation is possible due to the fact
that the solution to the master equation contains all the information on the
gauge structure of the theory. If a consistent interacting gauge theory can
be constructed, then the solution $S$ to the master equation associated with
the \textquotedblleft free\textquotedblright\ theory can be deformed into a
solution $\bar{S}$
\begin{equation}
S\rightarrow \bar{S}=S+gS_{1}+g^{2}S_{2}+g^{3}S_{3}+g^{4}S_{4}+\cdots ,\quad
\varepsilon (\bar{S})=0,\quad \mathrm{gh}(\bar{S})=0  \label{14h}
\end{equation}%
of the master equation for the deformed theory that displays the same ghost
and antifield spectra, namely,
\begin{equation}
(\bar{S},\bar{S})=0.  \label{15h}
\end{equation}%
According to the deformation parameter $g$, equation (\ref{15h}) splits
into:
\begin{align}
g^{0}& :(S,S)=0,  \label{34} \\
g^{1}& :sS_{1}=0,  \label{35} \\
g^{2}& :\tfrac{1}{2}(S_{1},S_{1})+sS_{2}=0,  \label{36} \\
g^{3}& :(S_{1},S_{2})+sS_{3}=0,  \label{37} \\
g^{4}& :\tfrac{1}{2}(S_{2},S_{2})+(S_{1},S_{3})+sS_{4}=0,  \label{38} \\
& \vdots  \notag
\end{align}%
The first equation is satisfied by hypothesis. The remaining ones are to be
solved recursively, from lower to higher orders, such that each equation
corresponding to a given order of perturbation theory, say $i$ ($i\geq 1$),
contains a single unknown functional, namely, the deformation of order $i$, $%
S_{i}$. Once the deformation equations (\ref{35})--(\ref{38}), etc., have
been solved by means of specific cohomological techniques, from the
consistent nontrivial deformed solution to the master equation one can
extract all the information on the gauge structure of the resulting
interacting theory. It is important to mention that the antifield-BRST
deformation method briefly exposed in the above has been successfully
applied to various models~\cite%
{YM1993,FT1996,NPB2001,Ikeda1,Weyl,Chiral,Exotic,SUGRA,PRD2003,Ikeda2,bf4,bf2,bf7,bf3}%
.

\section{Consistent interactions between a collection of scalar fields and
one vector field: deformed solution to the master equation \label{master}}

In the sequel we apply the deformation procedure exposed previously with the
purpose of generating consistent interacting gauge theories in $D=4$ whose
free limit is precisely the gauge theory described by relations (\ref{2})
and (\ref{2h}). We are interested only in (nontrivial) deformations that
comply with the standard hypotheses from field theory: analyticity in the
coupling constant, Lorentz covariance, spacetime locality, and Poincar\'{e}
invariance. Moreover, we require that the maximum number of derivatives
allowed within the interaction vertices is equal to two, i.e. the maximum
number of derivatives from the free Lagrangian (derivative-order assumption).

If we make the notation $S_{1}=\int d^{4}x\,a$, with $a$ a local function,
then equation (\ref{35}), which we have seen that controls the first-order
deformation, takes the local form
\begin{equation}
sa=\partial _{\mu }j^{\mu },\qquad \mathrm{gh}(a)=0,\qquad \varepsilon (a)=0,
\label{3.1}
\end{equation}
for some local $j^{\mu }$. Its solution is unique up to addition of trivial
quantities $a\rightarrow a^{\prime }=a+s\bar{a}+\partial _{\mu } \bar{j}%
^{\mu }$, $j^{\mu }\rightarrow j^{\prime \mu }=j^{\mu }+s\bar{j}^{\mu
}+\partial _{\nu }k^{\nu \mu }$ (with $k^{\nu \mu }=-k^{\mu \nu }$), in the
sense that $sa-\partial _{\mu }j^{\mu }\equiv sa^{\prime }-\partial _{\mu
}j^{\prime \mu }=0$. At the same time, if the general solution to (\ref{3.1}%
) is found to be completely trivial, $a=s\bar{a}+\partial _{\mu }\bar{j}%
^{\mu }$, then it can be made to vanish, $a=0$. In other words, $a$ is
constrained to pertain to a nontrivial class of the local BRST cohomology
(cohomology of $s$ modulo $d$ --- with $d$ the exterior differential in
spacetime) in $\mathrm{gh}=0$ computed in the algebra of (local)
nonintegrated densities, $H^{0}(s|d)$. In addition, all such solutions for $%
a $ will be selected such as to comply with the working hypotheses mentioned
in the above. The nonintegrated density of the first-order deformation
splits naturally into three components
\begin{equation}
a=a^{(A)}+a^{(\varphi )}+a^{\mathrm{int}},  \label{3.2}
\end{equation}
where $a^{(A)}$ and $a^{(\varphi )}$ describe the self-interactions of the
vector field $A^{\mu }$ and respectively of the scalar fields $\{\varphi
^{A}\}$, whereas $a^{\mathrm{int}}$ governs the couplings among them. The
three components display different contents of BRST generators ($a^{(A)}$
involves only $\{A^{\mu },\eta ,A_{\mu }^{\ast },\eta ^{\ast }\}$, $%
a^{(\varphi )}$ only $\{\varphi ^{A},\varphi _{A}^{\ast }\}$, and $a^{%
\mathrm{int}}$ mixes both sectors), such that equation (\ref{3.1}) becomes
equivalent to three independent equations, one for each piece,
\begin{equation}
sa^{(A)} =\partial _{\mu }j_{(A)}^{\mu }, \qquad sa^{(\varphi )} =\partial
_{\mu }j_{(\varphi )}^{\mu }, \qquad sa^{\mathrm{int}} =\partial _{\mu } j_{%
\mathrm{int}}^{\mu }.  \label{3.3}
\end{equation}
The solution to the first equation from (\ref{3.3}) is completely trivial
\cite{YM1993}, $a^{(A)}=0$, while the solution to the second equation
reduces to its component of antifield number $0$
\begin{equation}
a^{(\varphi )}=\tfrac{1}{2}\mu _{AB}(\varphi )\big(\partial _{\mu }\varphi
^{A}\big)\partial ^{\mu }\varphi ^{B}-\mathcal{V}(\varphi ),  \label{3.4}
\end{equation}
where $\mu _{AB}$ and $\mathcal{V}$ are some arbitrary, smooth real
functions depending only on the undifferentiated scalar fields, with
\begin{equation}
\mu _{AB}(\varphi )=\mu _{BA}(\varphi ), \qquad \mu _{AB}(\varphi )\neq
\frac{\partial u_{A}(\varphi )}{\partial \varphi ^{B}}+\frac{\partial
u_{B}(\varphi )}{\partial \varphi ^{A}}.  \label{3.5}
\end{equation}
Conditions (\ref{3.5}) ensure the nontriviality of $a^{(\varphi )}$ in $%
H^{0}(s|d)$.

In order to analyze the solutions to the last equation from (\ref{3.3}) we
decompose $a^{\mathrm{int}}$ along the antifield number. Since the starting
free theory is linear and its Cauchy order is equal to $2$ (see (\ref{2})
and (\ref{2h})), it follows that we can stop the previously mentioned
decomposition in antifield number $2$, $a^{\mathrm{int}}=a_{0}^{\mathrm{int}%
}+a_{1}^{\mathrm{int}}+a_{2}^{\mathrm{int}}$, with $\mathrm{agh}(a_{k}^{%
\mathrm{int}})=k$. Relying on the requirement $\mathrm{gh}(a)=0$, it results
that $\mathrm{pgh}(a_{k})=k$, and hence we have that $a_{2}^{\mathrm{int}}=%
\bar{\alpha}_{2}\eta ^{2}\equiv 0$ due to the fermionic behaviour of the
ghost $\eta $. In consequence, $a^{\mathrm{int}}$ reduces to the sum between
its first two components only, $a^{\mathrm{int}}=a_{0}^{\mathrm{int}}+a_{1}^{%
\mathrm{int}}$. Inserting this decomposition of $a^{\mathrm{int}}$ together
with splitting (\ref{3h}) of $s$ into the last equation from (\ref{3.3}), we
arrive at
\begin{align}
\gamma a_{1}^{\mathrm{int}}& =0,  \label{61} \\
\delta a_{1}^{\mathrm{int}}+\gamma a_{0}^{\mathrm{int}}& =\partial _{\mu }j_{%
\mathrm{int},0}^{\mu }.  \label{62}
\end{align}%
Strictly speaking, equation (\ref{61}) should have been written like $\gamma
a_{1}^{\mathrm{int}}=\partial _{\mu }j_{\mathrm{int},1}^{\mu }$. Since the
antifield number of both hand sides of this equation is strictly positive
(equal to $1$), it can be safely replaced by its homogeneous version without
loss of nontrivial terms, namely, one can always take $j_{\mathrm{int}%
,1}^{\mu }=0$. The proof of this result is done in a standard manner (for
instance, see~\cite%
{CMP1995b,NPB2001,PRD2003,IJMPA2004,IJGMMP2004,JHEP2006,PRD2006}). Equation (%
\ref{61}) shows that $a_{1}^{\mathrm{int}}$ can be taken as a $\gamma $%
-closed object of pure ghost number one. By means of formulas (\ref{4h})--(%
\ref{7h}), (\ref{10h}), and (\ref{11h}), we find that
\begin{equation}
a_{1}^{\mathrm{int}}=\big(A_{\mu }^{\ast }h^{\mu }([\varphi ],[F_{\mu \nu
}])+\varphi _{A}^{\ast }h^{A}([\varphi ],[F_{\mu \nu }])\big)\eta ,
\label{3.6}
\end{equation}
where the notation $h([y])$ means that $h$ depends on $y$ and its spacetime
derivatives up to a finite order. The existence of the solution $a_{0}^{%
\mathrm{int}}$ to equation (\ref{62}) requires that $\alpha _{1}=A_{\mu
}^{\ast }h^{\mu }([\varphi ],[F_{\mu \nu }])+\varphi _{A}^{\ast
}h^{A}([\varphi ],[F_{\mu \nu }])$ should be a nontrivial element of the
local homology of the Koszul--Tate differential in $\mathrm{agh}=1$, $%
H_{1}(\delta |d)$ (meaning that $\delta \alpha _{1}=\partial _{\mu }l^{\mu }$%
, with $\alpha _{1}\neq \delta \beta _{2} +\partial _{\mu } k^{\mu }$).
Taking into account the working hypotheses (including the derivative-order
assumption), after some computation we infer that the most general
nontrivial representative of $H_{1}(\delta |d)$ corresponds to
\begin{equation}
h^{\mu }([\varphi ],[F_{\mu \nu }])=0,\qquad h^{A}([\varphi ],[F_{\mu \nu
}])=T^{AB}k_{BC}\varphi ^{C}+n^{A},  \label{3.7}
\end{equation}
where $T^{AB}$ and $n^{A}$ are some arbitrary, real constants, with
\begin{equation}
T^{AB}=-T^{BA}.  \label{TAB}
\end{equation}
Then, from (\ref{3.6}) and (\ref{3.7}) we obtain that
\begin{equation}
a_{1}^{\mathrm{int}}=\varphi _{A}^{\ast }\big(T^{AB}k_{BC}\varphi ^{C}+n^{A} %
\big)\eta .  \label{3.8}
\end{equation}
Substituting (\ref{3.8}) in (\ref{62}) we deduce the component of antifield
number $0$
\begin{equation}
a_{0}^{\mathrm{int}}= -k_{AB}\big(T_{\hphantom{A}C}^{A}\varphi ^{C}+n^{A}%
\big)A_{\mu }\partial ^{\mu }\varphi ^{B} +\tfrac{1}{2}F_{\mu \nu } \big(%
U(\varphi )F^{\mu \nu }+\varepsilon ^{\mu \nu \rho \lambda }\tilde{U}%
(\varphi )F_{\rho \lambda }\big),  \label{3.9}
\end{equation}
with $T_{\hphantom{A}C}^{A}=T^{AB}k_{BC}$. In formula (\ref{3.9}) the
objects $U$ and $\tilde{U}$ denote some arbitrary, smooth real functions
depending on the undifferentiated scalar fields and $\varepsilon ^{\mu \nu
\rho \lambda }$ stand for the components of the Levi-Civita symbol in $D=4$.
In order to avoid trivial couplings the two functions $U$ and $\tilde{U}$
should contain no additive constants. In consequence, the first-order
deformation of the solution to the master equation reads
\begin{equation}
S_{1}=\int d^{4}x\big(a^{(\varphi )}+a_{1}^{\mathrm{int}}+a_{0}^{\mathrm{int}%
}\big),  \label{3.9x}
\end{equation}
with $a^{(\varphi )}$, $a_{1}^{\mathrm{int}}$, and $a_{0}^{\mathrm{int}}$
governed by relations (\ref{3.4})--(\ref{3.5}), (\ref{3.8}), and (\ref{3.9}%
), respectively.

Next, we investigate equation (\ref{36}). By direct computation, we arrive
at
\begin{align}
\tfrac{1}{2}(S_{1},S_{1})=& s\bigg\{ \int d^{4}x \Big[ \mu _{AB}(\varphi
)\partial ^{\mu }\varphi ^{A} -\tfrac{1}{2}k_{AB}A^{\mu } \big( T_{%
\hphantom{A}C}^{A} \varphi ^{C} +n^{A}\big) \Big] \big( T_{\hphantom{B}%
D}^{B} \varphi ^{D}  \notag \\
& +n^{B} \big) A_{\mu } \bigg\}+\int d^{4}x \bigg\{ -\frac{\partial \mathcal{%
V}(\varphi )}{\partial \varphi ^{A}} \big( T_{\hphantom{A}C}^{A} \varphi
^{C} +n^{A} \big) \eta  \notag \\
& +\tfrac{1}{2}\frac{\partial U(\varphi )}{\partial \varphi ^{A}}\big( T_{%
\hphantom{A}C}^{A}\varphi ^{C} +n^{A} \big) F_{\mu \nu } F^{\mu \nu } \eta
\notag \\
& +\tfrac{1}{2}\frac{\partial \tilde{U}(\varphi )}{\partial \varphi ^{A}} %
\big( T_{\hphantom{A}C}^{A} \varphi ^{C} +n^{A} \big) \varepsilon ^{\mu \nu
\rho \lambda } F_{\mu \nu } F_{\rho \lambda } \eta  \notag \\
& +\tfrac{1}{2} \bigg[ \mu _{AC}(\varphi )T_{\hphantom{C}B}^{C} +\mu
_{BC}(\varphi )T_{\hphantom{C}A}^{C}  \notag \\
& +\frac{\partial \mu _{AB}(\varphi )}{\partial \varphi ^{C}} \big( T_{%
\hphantom{C}D}^{C} \varphi ^{D} +n^{C}\big)\bigg]\big( \partial _{\mu }
\varphi ^{A}\big) \big(\partial ^{\mu }\varphi ^{B}\big) \eta \bigg\}.
\label{3.10}
\end{align}
Formulas (\ref{36}) and (\ref{3.10}) show that the first-order deformation
is consistent at order $g^{2}$ if and only if the following relations are
fulfilled:
\begin{align}
\frac{\partial \mathcal{V}(\varphi )}{\partial \varphi ^{A}} \big( T_{%
\hphantom{A}B}^{A} \varphi ^{B} +n^{A} \big) & =0,  \label{192} \\
\frac{\partial U(\varphi )}{\partial \varphi ^{A}} \big( T_{\hphantom{A}%
B}^{A} \varphi ^{B} +n^{A} \big) & =0,  \label{192x} \\
\frac{\partial \tilde{U}(\varphi )}{\partial \varphi ^{A}} \big( T_{%
\hphantom{A}B}^{A} \varphi ^{B} +n^{A} \big) & =0,  \label{192y} \\
\mu _{AC}(\varphi )T_{\hphantom{C}B}^{C} +\mu _{BC}(\varphi )T_{\hphantom{C}%
A}^{C} +\frac{\partial \mu _{AB}(\varphi )}{\partial \varphi ^{C}} \big( T_{%
\hphantom{C}D}^{C} \varphi ^{D} +n^{C} \big) & =0.  \label{193}
\end{align}
In what follows we call (\ref{192})--(\ref{193}) consistency equations.
Under these circumstances, from (\ref{3.10}) we find that
\begin{equation}
S_{2} =\int d^{4}x \Big[ -\mu _{AB}(\varphi ) \partial ^{\mu }\varphi ^{A} +
\tfrac{1}{2} k_{AB} A^{\mu } \big( T_{\hphantom{A}C}^{A} \varphi ^{C}+n^{A} %
\big) \Big] \big( T_{\hphantom{B}D}^{B}\varphi ^{D} +n^{B}\big)A_{\mu }.
\label{3.11}
\end{equation}
With the help of relations (\ref{3.9x}) and (\ref{3.11}) we compute the
antibracket $(S_{1},S_{2})$ and then, by means of equation (\ref{37}), we
deduce the third-order deformation
\begin{equation}
S_{3}= \int d^{4}x \Big[ \tfrac{1}{2} \mu _{AB}(\varphi ) \big( T_{%
\hphantom{A}C}^{A} \varphi ^{C} +n^{A} \big) \big( T_{\hphantom{B}D}^{B}
\varphi ^{D} +n^{B} \big) A_{\mu }A^{\mu }\Big].  \label{3.12}
\end{equation}
Simple computation provides $(S_{2},S_{2}) =0$ and $(S_{1},S_{3})=0$, so the
solution to equation (\ref{38}) can be taken as $S_{4}=0$. Then, all the
remaining higher-order deformations can be chosen to vanish: $S_{k}=0$, $k>4$%
.

In consequence, we can state that the complete deformed solution to the
master equation for the model under study, which is consistent to all orders
in the coupling constant, reads
\begin{equation}
\bar{S}=S+gS_{1}+g^{2}S_{2}+g^{3}S_{3},  \label{defsolmast}
\end{equation}
where $S$, $S_{1}$, $S_{2}$, and $S_{3}$ are given by formulas (\ref{13h}), (%
\ref{3.9x}), (\ref{3.11}), and (\ref{3.12}), respectively. The fully
deformed solution to the master equations depends on two kinds of real
constants ($T^{AB}=-T^{BA}$ and $n^{A}$) and four types of smooth, real
functions of the undifferentiated scalar fields ($\mathcal{V}$, $U$, $\tilde{%
U}$, and $\mu _{AB}=\mu _{BA}$). In addition, the above constants and
functions are subject to the consistency equations (\ref{192})--(\ref{193}).
Thus, our procedure is consistent provided these equations possess solutions.

Everywhere in the sequel we assume that
\begin{equation}
\mathrm{rank}(T^{AB})\neq 0,\qquad A,B=\overline{1,N_{0}},\qquad N_{0}>1.
\label{condneq1}
\end{equation}
For the sake of generality, we consider that the matrix $T^{AB}$ may possess
some nontrivial null vectors
\begin{equation}
T_{\hphantom{A}B}^{A}\tau _{\hphantom{B}i}^{B}=0,\qquad i=\overline{1,N_{0}-%
\mathrm{rank}(T^{AB})}.  \label{modif1}
\end{equation}
It is understood that if $\mathrm{rank}(T^{AB})=N_{0}$, then relations (\ref%
{modif1}) are absent. Introducing the quantities
\begin{equation}
\Omega _{i}=k_{AB}\varphi ^{A}\tau _{\hphantom{B}i}^{B},\qquad
q_{i}=k_{AB}n^{A}\tau _{\hphantom{B}i}^{B},  \label{modif2}
\end{equation}
we find that a class of solutions to equations (\ref{192})--(\ref{193}) is
given by
\begin{align}
\mathcal{V}(\varphi )& =\mathcal{V}(r,\bar{\Omega}_{i},r_{\alpha }), &
U(\varphi )& =\vartheta (\bar{r},\bar{\Omega}_{i},\bar{r}_{\alpha }),
\label{3.13a} \\
\tilde{U}(\varphi )& =\varkappa (\bar{r},\bar{\Omega}_{i},\bar{r}_{\alpha }),
& \mu _{AB}(\varphi )& =k_{AB}\omega (\bar{r},\bar{\Omega}_{i},\bar{r}
_{\alpha }),  \label{3.13b}
\end{align}
where $r$, $\bar{r}$, and $\bar{\Omega}_{i}$ read
\begin{align}
r& =\tfrac{1}{2}k_{AB}\big(T_{\hphantom{A}C}^{A}\varphi ^{C}+n^{A}\big)\big(%
T_{\hphantom{B}D}^{B}\varphi ^{D}+n^{B}\big),  \label{3.14} \\
\bar{r}& =k_{AB}T_{\hphantom{A}C}^{A}\varphi ^{C}\big(\tfrac{1}{2}T_{%
\hphantom{B}D}^{B}\varphi ^{D}+n^{B}\big),  \label{3.15} \\
\bar{\Omega}_{i}& =\delta _{0q_{i}}\Omega _{i}\quad (\mathrm{%
no~summation~over}~i),  \label{3.15a}
\end{align}
while $r_{\alpha }(\varphi )$ are other solutions to the equations
\begin{equation}
\frac{\partial r_{\alpha }}{\partial \varphi ^{A}}\big(T_{\hphantom{A}%
B}^{A}\varphi ^{B}+n^{A}\big)=0,\qquad \alpha =1,\ldots  \label{ra}
\end{equation}
(if any) and $\bar{r}_{\alpha }(\varphi )$ are given by
\begin{equation}
\bar{r}_{\alpha }=r_{\alpha }-r_{\alpha }(\varphi ^{A}=0).  \label{ra1}
\end{equation}
In (\ref{3.13a}) and (\ref{3.13b}) $\mathcal{V}(r,\bar{\Omega}_{i},r_{\alpha
})$, $\vartheta (\bar{r},\bar{\Omega}_{i},\bar{r}_{\alpha })$, $\varkappa (%
\bar{r},\bar{\Omega}_{i},\bar{r}_{\alpha })$, and $\omega (\bar{r},\bar{%
\Omega}_{i},\bar{r}_{\alpha })$ are some arbitrary, smooth real functions of
their arguments and, in addition, $\vartheta (\bar{r},\bar{\Omega}_{i},\bar{r%
}_{\alpha })$, $\varkappa (\bar{r}, \bar{\Omega}_{i}, \bar{r}_{\alpha })$,
and $\omega (\bar{r},\bar{\Omega}_{i},\bar{r}_{\alpha })$ are constrained to
satisfy the conditions
\begin{equation}
\vartheta (0,0,0)=\varkappa (0,0,0)=\omega (0,0,0)=0.  \label{3.16}
\end{equation}
The above conditions ensure that the three functions denoted by $\vartheta (%
\bar{r},\bar{\Omega}_{i},\bar{r}_{\alpha })$, $\varkappa (\bar{r}, \bar{%
\Omega}_{i}, \bar{r}_{\alpha })$, and $\omega (\bar{r},\bar{\Omega}_{i},\bar{%
r}_{\alpha })$ contain no additive constants and, as a consequence, none of
the functions $U$, $\tilde{U}$, or $\mu _{AB}$ may exhibit trivial
components. We believe that relations (\ref{3.13a})--(\ref{3.16}) provide
the most general class of solutions to equations (\ref{192})--(\ref{193}),
but we do not insist on this matter. We remark that in the context of the
above solutions the constants $n^{A}$ remain arbitrary. In view of this, we
choose them such that
\begin{equation}
k_{AB}n^{A}n^{B}\neq 0.  \label{condneq2}
\end{equation}
The first formula from (\ref{modif2}) and relation (\ref{3.15a}) show that
the dependence on $\bar{\Omega}_{i}$'s in (\ref{3.13a}) and (\ref{3.13b})
may appear only in the presence of some nontrivial vectors $\tau _{%
\hphantom{B}i}^{B}$ that obey relations (\ref{modif1}). However, for a given
set of constants $n^{A}$ that fulfills (\ref{condneq2}), the presence of the
Kronecker delta $\delta _{0q_{i}}$ in (\ref{3.15a}) signalizes that the
dependence on $\bar{\Omega}_{i}$'s in (\ref{3.13a}) and (\ref{3.13b}) is
nontrivial if and only if the null vectors $\tau _{\hphantom{B}i}^{B}$
satisfy the conditions
\begin{equation}
q_{i}\equiv k_{AB}n^{A}\tau _{\hphantom{B}i}^{B}=0  \label{iffnon}
\end{equation}
for at least one $i\in \overline{1,N_{0}-\mathrm{rank}(T^{AB})}$.

Inserting relations (\ref{3.13a}) and (\ref{3.13b}) into (\ref{defsolmast}),
we obtain the final form of the deformed solution to the master equation
that is consistent to all orders in the coupling constant,
\begin{align}
\bar{S}=\int d^{4}x\Big[& -\tfrac{1}{4}F_{\mu \nu }F^{\mu \nu }+\tfrac{1}{2}%
g^{2}k_{AB}n^{A}n^{B}A_{\mu }A^{\mu } -g\mathcal{V}(r,\bar{\Omega}%
_{i},r_{\alpha })  \notag \\
& +\tfrac{1}{2}k_{AB}(1+g\omega (\bar{r},\bar{\Omega}_{i},\bar{r}_{\alpha }))%
\big(D_{\mu }\varphi ^{A}-2gn^{A}A_{\mu }\big)D^{\mu }\varphi ^{B}  \notag \\
& +\tfrac{1}{2}gF_{\mu \nu }\big(\vartheta (\bar{r},\bar{\Omega}_{i},\bar{r}%
_{\alpha })F^{\mu \nu }+\varepsilon ^{\mu \nu \rho \lambda }\varkappa (\bar{r%
},\bar{\Omega}_{i},\bar{r}_{\alpha })F_{\rho \lambda }\big)  \notag \\
& +\tfrac{1}{2}g^{3}k_{AB}\omega (\bar{r},\bar{\Omega}_{i},\bar{r}_{\alpha
}) n^{A}n^{B}A_{\mu }A^{\mu }  \notag \\
& +A_{\mu }^{\ast }\partial ^{\mu }\eta +g\varphi _{A}^{\ast }\big(%
T^{AB}k_{BC}\varphi ^{C}+n^{A}\big)\eta \Big],  \label{defsolmast1}
\end{align}
with the covariant derivative of the matter fields defined by
\begin{equation}
D_{\mu }\varphi ^{A}=\partial _{\mu }\varphi ^{A}-gT_{\hphantom{A}%
B}^{A}\varphi ^{B}A_{\mu }.  \label{dercov}
\end{equation}
The functional $\bar{S}$ satisfies by construction the equation
\begin{equation}
(\bar{S},\bar{S})=0.  \label{Sdef}
\end{equation}
Formulas (\ref{3.13a})--(\ref{condneq2}), (\ref{defsolmast1}), and (\ref%
{Sdef}) stand for the general results of the deformation procedure under the
current working hypotheses.

\section{Lagrangian formulation of emerging interacting gauge theories.
Gauge-fixed action\label{inter}}

Under these circumstances, from (\ref{defsolmast1}) and (\ref{3.13a})--(\ref%
{condneq2}) we extract all the ingredients correlated with the Lagrangian
formulation of the resulting interacting gauge theory. The antifield number $%
0$ piece in the deformed solution (\ref{defsolmast1}) is nothing but the
Lagrangian action of the emerging class of interacting gauge theories
\begin{align}
\bar{S}_{0}^{\mathrm{L}}\big[A^{\mu },\varphi ^{A}\big]=\int d^{4}x\Big[& -%
\tfrac{1}{4}F_{\mu \nu }F^{\mu \nu }+\tfrac{1}{2}g^{2}k_{AB}n^{A}n^{B}A_{\mu
}A^{\mu } -g\mathcal{V}(r,\bar{\Omega}_{i},r_{\alpha })  \notag \\
& +\tfrac{1}{2}k_{AB}(1+g\omega (\bar{r},\bar{\Omega}_{i},\bar{r}_{\alpha }))%
\big(D_{\mu }\varphi ^{A}-2gn^{A}A_{\mu }\big)D^{\mu }\varphi ^{B}  \notag \\
& +\tfrac{1}{2}gF_{\mu \nu }\big(\vartheta (\bar{r},\bar{\Omega}_{i},\bar{r}%
_{\alpha })F^{\mu \nu }+\varepsilon ^{\mu \nu \rho \lambda }\varkappa (\bar{r%
},\bar{\Omega}_{i},\bar{r}_{\alpha })F_{\rho \lambda }\big)  \notag \\
& +\tfrac{1}{2}g^{3}k_{AB}\omega (\bar{r},\bar{\Omega}_{i},\bar{r}_{\alpha
})n^{A}n^{B}A_{\mu }A^{\mu }\Big].  \label{3.17}
\end{align}
From the terms of antifield number $1$ present in (\ref{defsolmast1}) we
read the deformed gauge transformations (which leave invariant action (\ref%
{3.17})), namely,
\begin{equation}
\bar{\delta}_{\epsilon }A^{\mu }=\partial ^{\mu }\epsilon ,\qquad \bar{\delta%
}_{\epsilon }\varphi ^{A}=g\big(T_{\hphantom{A}B}^{A}\varphi ^{B} +n^{A}\big)%
\epsilon ,\quad A=\overline{1,N_{0}}.  \label{3.18}
\end{equation}
The previous gauge transformations are Abelian and irreducible. Relations (%
\ref{3.17}) and (\ref{3.18}) serve as the general output of steps (i) and
(ii) discussed in the introductory section. Now, we are in the position to
emphasize and detail the main results announced in introduction.

It is well known that the deformation procedure does not change the number
of physical degrees of freedom of the starting theory \cite{PLB1993,CM1998}.
Due to the fact that the matrix of elements $k_{AB}$ was taken by assumption
to be positively defined and the constants $n^{A}$ satisfy condition (\ref%
{condneq2}), we find that $k_{AB}n^{A}n^{B}>0$. In consequence, the object
\begin{equation}
\tfrac{1}{2}g^{2}k_{AB}n^{A}n^{B}A_{\mu }A^{\mu }\equiv \tfrac{1}{2}%
M_{(A)}^{2}A_{\mu }A^{\mu }  \label{3.19}
\end{equation}
from (\ref{3.17}) is precisely a mass term for the vector field $A_{\mu }$.
It is clear that the quantity $(1/2)g^{3}k_{AB}\omega (\bar{r},\bar{\Omega}%
_{i},\bar{r}_{\alpha })n^{A}n^{B}A_{\mu }A^{\mu }$ cannot generate mass for $%
A_{\mu }$ due to the fact that $\omega (\bar{r},\bar{\Omega}_{i},\bar{r}%
_{\alpha })$ contains no additive constants (see requirement (\ref{3.16})).
Then, the vector field present in (\ref{3.17}) possesses precisely three
physical degrees of freedom. It is easy to see that the mass term (\ref{3.19}%
) exists irrespective of the number of scalar fields from the collection.
Meanwhile, we remark that the term $-gk_{AB}A_{\mu }n^{A}\partial ^{\mu
}\varphi ^{B}$ from (\ref{3.17}) is non-propagating. As a result, the linear
combination of scalar fields $\phi \equiv k_{AB}n^{A}\varphi ^{B}$
represents an unphysical degree of freedom, so there are $(N_{0}-1)$ scalar
physical degrees of freedom in (\ref{3.17}). Therefore, the deformed action (%
\ref{3.17}) describes a system with $(N_{0}+2)$ physical degrees of freedom,
like its free limit (\ref{2}). We observe that the mass term (\ref{3.19}) is
generated by the nonvanishing arbitrary constants $n^{A}$. On the one hand,
the existence of the constants $n^{A}$ in (\ref{3.17}) is a consequence of
the existence of the one-parameter global symmetry $\Delta _{\theta }\varphi
^{A}=\big(T^{AB}k_{BC}\varphi ^{C}+n^{A}\big)\theta $ of (the free) action (%
\ref{2}). Thus, the appearance of the mass term (\ref{3.19}) is a direct
consequence of the deformation method employed here in the context of the
free limit described by action (\ref{2}). At the same time, the constants $%
n^{A}$ are involved also in the deformed gauge transformations of the scalar
fields from (\ref{3.18}), which are nothing but the gauge versions of the
one-parameter global transformations mentioned previously. On the other
hand, none of the functions $\mathcal{V}(r,\bar{\Omega}_{i},r_{\alpha })$, $%
\omega (\bar{r},\bar{\Omega}_{i},\bar{r}_{\alpha })$, $\vartheta (\bar{r},%
\bar{\Omega}_{i},\bar{r}_{\alpha })$, or $\varkappa (\bar{r},\bar{\Omega}%
_{i},\bar{r}_{\alpha })$ that parameterize action (\ref{3.17}) may
contribute to the mass of the vector field. Actually, these functions are
involved in (\ref{3.17}) as follows: (A) $\mathcal{V}(r, \bar{\Omega}_{i},
r_{\alpha })$ describes the derivative-free self-interactions and possibly
some mass terms of the scalar fields; (B) $\omega (\bar{r},\bar{\Omega}_{i},%
\bar{r}_{\alpha })$ controls the self-interactions of the form $\omega (\bar{%
r},\bar{\Omega}_{i},\bar{r}_{\alpha })k_{AB}\big(\partial _{\mu }\varphi ^{A}%
\big)\partial ^{\mu }\varphi ^{B}$ among the scalar fields as well as some
cross-couplings between the vector field and the matter sector; (C) $%
\vartheta (\bar{r},\bar{\Omega}_{i},\bar{r}_{\alpha })$ and $\varkappa (\bar{%
r}, \bar{\Omega}_{i}, \bar{r}_{\alpha })$ are responsible solely for some
cross-couplings between the Abelian gauge field and the matter scalars.
Until now we proved that the procedure based on steps (i) and (ii) leads to
results (iii) and (iv) announced in Section \ref{intro}.

In order to argue that result (v) from Section \ref{intro} also holds, we
need to construct the gauge-fixed action corresponding to the deformed
solution of the master equation given in (\ref{defsolmast1}). In view of
this, we introduce the cohomologically trivial pairs $\{B,B^{\ast }\}$ and $%
\{\bar{\eta},\bar{\eta}^{\ast }\}$, with $\mathrm{gh}(B)=0=\mathrm{gh}(\bar{%
\eta}^{\ast })$, $\mathrm{gh}(B^{\ast })=-1=\mathrm{gh}(\bar{\eta})$, and $%
\varepsilon (B)=0=\varepsilon (\bar{\eta}^{\ast })$, $\varepsilon (B^{\ast
})=1=\varepsilon (\bar{\eta})$. Consequently, the non-minimal solution to
the master equation corresponding to (\ref{defsolmast1}) is given by $\bar{S}%
_{\mathrm{nm}}=\bar{S}+\int d^{4}x\,\bar{\eta}^{\ast }B$. Since we have
already identified the unphysical scalar degree of freedom, it is no longer
necessary to enforce the unitary gauge. Instead, we work with the $R_{\xi }$
gauge implemented via the gauge-fixing fermion
\begin{equation}
K=-\int d^{4}x\bar{\eta}\Big(\partial _{\mu }A^{\mu }+\xi gk_{AB}\varphi
^{A}n^{B}-\tfrac{1}{2}\xi B\Big),  \label{3.20}
\end{equation}%
where $\xi $ is an arbitrary real constant. As a result, the gauge-fixed
action becomes $\bar{S}_{K}=\bar{S}_{\mathrm{nm}}\Big[\Phi ^{\alpha
_{0}},\Phi _{\alpha _{0}}^{\ast }=\frac{\delta K}{\delta \Phi ^{\alpha _{0}}}%
\Big]$, where $\Phi ^{\alpha _{0}}$ is a collective notation for all the
fields/ghosts. If we eliminate the auxiliary field $B$ from $\bar{S}_{K}$
according to its field equation, we infer that
\begin{align}
\bar{S}_{K}=\int d^{4}x\Big[& -\tfrac{1}{4}F_{\mu \nu }F^{\mu \nu }+\tfrac{1%
}{2}g^{2}k_{AB}n^{A}n^{B}A_{\mu }A^{\mu }-\tfrac{1}{2\xi }(\partial _{\mu
}A^{\mu })^{2}  \notag \\
& +\tfrac{1}{2}k_{AB}\big(D_{\mu }\varphi ^{A}\big)D^{\mu }\varphi ^{B}-g%
\mathcal{V}(r,\bar{\Omega}_{i},r_{\alpha })  \notag \\
& +\tfrac{1}{2}gk_{AB}\omega (\bar{r},\bar{\Omega}_{i},\bar{r}_{\alpha })%
\big[\big(D_{\mu }\varphi ^{A}-2gn^{A}A_{\mu }\big)D^{\mu }\varphi
^{B}+g^{2}n^{A}n^{B}A_{\mu }A^{\mu }\big]  \notag \\
& +\tfrac{1}{2}gF_{\mu \nu }\big(\vartheta (\bar{r},\bar{\Omega}_{i},\bar{r}%
_{\alpha })F^{\mu \nu }+\varepsilon ^{\mu \nu \rho \lambda }\varkappa (\bar{r%
},\bar{\Omega}_{i},\bar{r}_{\alpha })F_{\rho \lambda }\big)  \notag \\
& -\tfrac{1}{2}\xi g^{2}k_{AC}k_{BD}n^{C}n^{D}\varphi ^{A}\varphi
^{B}+g^{2}k_{AB}n^{A}T_{\hphantom{B}C}^{B}\varphi ^{C}A_{\mu }A^{\mu }
\notag \\
& +(\partial _{\mu }\bar{\eta})\partial ^{\mu }\eta -\xi g^{2}\bar{\eta}%
k_{AB}n^{B}\big(T_{\hphantom{A}C}^{A}\varphi ^{C}+n^{A}\big)\eta \Big].
\label{3.21}
\end{align}%
The gauge-fixed action (\ref{3.21}) is invariant under the gauge-fixed BRST
transformations
\begin{align}
\bar{s}_{K}A^{\mu }& =\partial ^{\mu }\eta , & \bar{s}_{K}\varphi ^{A}& =g%
\big(T_{\hphantom{A}C}^{A}\varphi ^{C}+n^{A}\big)\eta ,  \label{3.21a} \\
\bar{s}_{K}\eta & =0, & \bar{s}_{K}\bar{\eta}& =\frac{1}{\xi }\big(\partial
_{\mu }A^{\mu }+\xi gk_{AB}\varphi ^{A}n^{B}\big).  \label{3.21b}
\end{align}%
Formula (\ref{3.21}) emphasizes the following features: (I) the massive
vector field propagator behaves like $\tilde{\Delta}_{\mathrm{F}\mu \nu }(%
\bar{p})\sim |\bar{p}|^{-2}$ for large Euclidean momenta $|\bar{p}%
|\rightarrow \infty $, just like in the massless case; (II) the unphysical
degrees of freedom $\phi \equiv k_{AB}n^{A}\varphi ^{B}$ and $\{\bar{\eta}%
,\eta \}$ acquire mass; (III) the scalar fields may be coupled nontrivially
to the ghosts. Conclusion (I) is nothing but result (v). Clearly, the last
conclusion highlights a propagator behaviour that is different from the
purely Proca case exposed in Ref. \cite{lavinia}.

In agreement with the previous discussion regarding the properties of the
coupled gauge model (see statements (B) and (C) from the previous
paragraph), we notice that the functions $\omega (\bar{r},\bar{\Omega}_{i},%
\bar{r}_{\alpha })$, $\vartheta (\bar{r}, \bar{\Omega}_{i}, \bar{r}_{\alpha
})$, and $\varkappa (\bar{r},\bar{\Omega}_{i},\bar{r}_{\alpha })$ are less
relevant. For the sake of simplicity, we will set them equal to zero in what
follows
\begin{equation}
\omega (\bar{r},\bar{\Omega}_{i},\bar{r}_{\alpha })=0,\qquad \vartheta (\bar{%
r},\bar{\Omega}_{i},\bar{r}_{\alpha })=0,\qquad \varkappa (\bar{r},\bar{%
\Omega}_{i},\bar{r}_{\alpha })=0.  \label{3.19a}
\end{equation}
All the above results remain valid in the presence of (\ref{3.19a}) since
the functions $\omega (\bar{r},\bar{\Omega}_{i},\bar{r}_{\alpha })$, $%
\vartheta (\bar{r},\bar{\Omega}_{i},\bar{r}_{\alpha })$, and $\varkappa (%
\bar{r},\bar{\Omega}_{i},\bar{r}_{\alpha })$ were so far arbitrary.

We remark that all the outcomes obtained until now are entirely independent
of the Higgs mechanism.

\section{Comparison with the Abelian Higgs mechanism\label{Higgs}}

Initially, we briefly address the Abelian Higgs mechanism in the presence of
a collection of $N_{0}$ real scalar fields. In this situation the starting
point is given by the action (we recall that the covariant derivative $%
D_{\mu }\varphi ^{A}$ is introduced in (\ref{dercov}))%
\begin{align}
\bar{S}_{0_{T,\mathrm{Higgs}}}^{\mathrm{L}}\big[A^{\mu },\varphi ^{A}\big]%
=\int d^{4}x\Big[& -\tfrac{1}{4}F_{\mu \nu }F^{\mu \nu }+\tfrac{1}{2}k_{AB}%
\big(D_{\mu }\varphi ^{A}\big)D^{\mu }\varphi ^{B}  \notag \\
& -V_{1}^{\mathrm{Higgs}}(\varphi ^{A})\Big],  \label{compX}
\end{align}%
which is assumed to be invariant under the Abelian and irreducible gauge
transformations
\begin{equation}
\delta _{\epsilon }^{\prime }A^{\mu }=\partial ^{\mu }\epsilon ,\qquad
\delta _{\epsilon }^{\prime }\varphi ^{A}=gT_{\hphantom{A}B}^{A}\varphi
^{B}\epsilon .  \label{compX1}
\end{equation}%
Formulas (\ref{compX}) and (\ref{compX1}) are nothing but a generalization
of relations (\ref{1H}) and (\ref{3Hx}) for an arbitrary $N_{0}$. The gauge
invariance of (\ref{compX}) under (\ref{compX1}) is equivalent to the fact
that the function $V_{1}^{\mathrm{Higgs}}(\varphi ^{A})$ is gauge-invariant,
i.e.,
\begin{equation}
\frac{\partial V_{1}^{\mathrm{Higgs}}(\varphi ^{A})}{\partial \varphi ^{A}}%
T_{\hphantom{A}B}^{A}\varphi ^{B}=0.  \label{comp4}
\end{equation}%
In addition, we presume that the function $V_{1}^{\mathrm{Higgs}}(\varphi ^{A})$
possesses an absolute minimum for a (nonvanishing) constant scalar field
configuration
\begin{equation}
\varphi ^{A}=v_{0}^{A},  \label{higs1}
\end{equation}%
but make no other supplementary presumptions on $V_{1}^{\mathrm{%
Higgs}}$. Defining some new fields by
\begin{equation}
\tilde{\varphi}^{A}=\varphi ^{A}-v_{0}^{A},  \label{higs3}
\end{equation}%
whose associated field operators display zero vacuum expectation values, and
rewriting formulas (\ref{compX}) and (\ref{compX1}) in terms of (\ref{higs3}%
), we arrive at the action
\begin{align}
& \bar{S}_{0_{T,\mathrm{Higgs}}}^{\mathrm{L}}\big[A^{\mu },\tilde{\varphi}%
^{A}\big]=\int d^{4}x\Big[-\tfrac{1}{4}F_{\mu \nu }F^{\mu \nu }+\tfrac{1}{2}%
g^{2}k_{AB}T_{\hphantom{A}C}^{A}T_{\hphantom{B}D}^{B}v_{0}^{C}v_{0}^{D}A_{%
\mu }A^{\mu }  \notag \\
& +\tfrac{1}{2}k_{AB}\big(D_{\mu }\tilde{\varphi}^{A}-2gT_{\hphantom{A}%
C}^{A}v_{0}^{C}A_{\mu }\big)D^{\mu }\tilde{\varphi}^{B}-V_{1}^{\mathrm{Higgs}%
}(\tilde{\varphi}^{A}+v_{0}^{A})\Big],  \label{higs4}
\end{align}%
invariant under the gauge transformations
\begin{equation}
\delta _{\epsilon }^{\prime }A^{\mu }=\partial ^{\mu }\epsilon ,\qquad
\delta _{\epsilon }^{\prime }\tilde{\varphi}^{A}=gT_{\hphantom{A}B}^{A}\big(%
\tilde{\varphi}^{B}+v_{0}^{B}\big)\epsilon .  \label{higs5}
\end{equation}%
Relations (\ref{higs4}) and (\ref{higs5}) stand for the final output of the
Abelian Higgs mechanism in the presence of a collection of $N_{0}$ real
scalar fields and show that the vector field acquires the square mass $%
g^{2}k_{AB}T_{\hphantom{A}C}^{A}T_{\hphantom{B}D}^{B}v_{0}^{C}v_{0}^{D}$.
Formula (\ref{comp4}) written in terms of the transformed scalar fields (\ref%
{higs3}) is equivalent to the gauge-invariance of the function $V_{1}^{%
\mathrm{Higgs}}(\tilde{\varphi}^{A}+v_{0}^{A})$ under transformations (\ref%
{higs5})
\begin{equation}
\frac{\partial V_{1}^{\mathrm{Higgs}}(\tilde{\varphi}^{A}+v_{0}^{A})}{%
\partial \tilde{\varphi}^{A}}T_{\hphantom{A}B}^{A}\big(\tilde{\varphi}%
^{B}+v_{0}^{B}\big)=0.  \label{higs5.1}
\end{equation}%
The square masses of the scalar fields are the eigenvalues of the mass
matrix $m_{AB}=\left. \frac{\partial ^{2}V_{1}^{\mathrm{Higgs}}(\tilde{%
\varphi}^{A}+v_{0}^{A})}{\partial \tilde{\varphi}^{A}\partial \tilde{\varphi}%
^{B}}\right\vert _{\tilde{\varphi}^{A}=0}$. By differentiating (\ref{higs5.1}%
) with respect to $\tilde{\varphi}^{B}$, particularizing the resulting
formula to $\tilde{\varphi}^{A}=0$, and taking into account that $\left.
\frac{\partial V_{1}^{\mathrm{Higgs}}(\tilde{\varphi}^{A}+v_{0}^{A})}{%
\partial \tilde{\varphi}^{A}}\right\vert _{\tilde{\varphi}^{A}=0}=0$ ($%
\Leftrightarrow \left. \frac{\partial V_{1}^{\mathrm{Higgs}}(\varphi ^{A})}{%
\partial \varphi ^{A}}\right\vert _{\varphi ^{A}=v_{0}^{A}}=0$), we obtain
the relations $m_{AB}T_{\hphantom{A}C}^{A}v_{0}^{C}=0$, which show that the
maximum possible rank of the scalar mass matrix is equal to $(N_{0}-1)$.
This means that at least one scalar field (Goldstone mode) is massless. The
masses of the remaining scalars depend on the concrete form of $V_{1}^{%
\mathrm{Higgs}}(\varphi ^{A})$.

The above discussion allows us to conclude that: (a) the Higgs mechanism is
applicable if the gauge-invariant function $V_{1}^{\mathrm{Higgs}}(\varphi
^{A})$ that appears in (\ref{compX}) possesses an absolute minimum for a
nonvanishing scalar field configuration.

The starting point of our method is represented by the free limit given by
relations (\ref{2}) and (\ref{2h}). At the same time, the starting point of
the Higgs mechanism is provided by an interacting theory (see formulas (\ref%
{compX}) and (\ref{compX1})). Thus, in order to correctly compare our
procedure with the Higgs mechanism, it is necessary to consider an
appropriate starting point. In view of this, we begin with an interacting
Lagrangian action
\begin{equation}
\bar{S}_{0_{T}}^{\mathrm{L}}\big[A^{\mu },\varphi ^{A}\big]=\int d^{4}x\Big[-%
\tfrac{1}{4}F_{\mu \nu }F^{\mu \nu }+\tfrac{1}{2}k_{AB}\big(D_{\mu }\varphi
^{A}\big)D^{\mu }\varphi ^{B}-V_{1}(\varphi ^{A})\Big],  \label{comp1}
\end{equation}%
where $V_{1}(\varphi ^{A})$ is an arbitrary, smooth real function of its
arguments. Action (\ref{comp1}) is assumed to be invariant under the gauge
transformations
\begin{equation}
\delta _{\epsilon }^{\prime }A^{\mu }=\partial ^{\mu }\epsilon ,\qquad
\delta _{\epsilon }^{\prime }\varphi ^{A}=gT_{\hphantom{A}B}^{A}\varphi
^{B}\epsilon .  \label{comp3}
\end{equation}%
This implies that the function $V_{1}(\varphi ^{A})$ is gauge-invariant, $%
\delta _{\epsilon }^{\prime }V_{1}(\varphi ^{A})=0$. We make no further
assumption on the function $V_{1}(\varphi ^{A})$, so formulas (\ref{comp1})
and (\ref{comp3}) can be regarded like a more general starting point than
relations (\ref{compX}) and (\ref{compX1}).

Now, we prove that starting from action (\ref{comp1}) and gauge
transformations (\ref{comp3}) we can derive a gauge theory with a massive
vector field. The main results of the deformation procedure exposed in the
above, more precisely formulas (\ref{defsolmast1}) and (\ref{Sdef}), offer a
general manner of finding such a theory. The strategy goes as follows.
Initially, we construct the solution to the master equation associated with
the theory governed by (\ref{comp1}) and (\ref{comp3}). It reads
\begin{align}
\bar{S}_{T}=\int d^{4}x\Big[& -\tfrac{1}{4}F_{\mu \nu }F^{\mu \nu }+\tfrac{1%
}{2}k_{AB}\big(D_{\mu }\varphi ^{A}\big)D^{\mu }\varphi ^{B}-V_{1}(\varphi
^{A})  \notag \\
& +A_{\mu }^{\ast }\partial ^{\mu }\eta +g\varphi _{A}^{\ast }T_{\hphantom{A}%
B}^{A}\varphi ^{B}\eta \Big].  \label{comp8}
\end{align}%
Now, we introduce a (local) functional of fields, ghosts, and antifields,
defined by
\begin{align}
\bar{S}_{n}=\int d^{4}x\Big(& \tfrac{1}{2}g^{2}k_{AB}n^{A}n^{B}A_{\mu
}A^{\mu }-gk_{AB}n^{A}A_{\mu }D^{\mu }\varphi ^{B}  \notag \\
& +V_{1}(\varphi ^{A})-g\mathcal{V}(r,\bar{\Omega}_{i},r_{\alpha })+g\varphi
_{A}^{\ast }n^{A}\eta \Big),  \label{comp12}
\end{align}%
where the arbitrary constants $n^{A}$ still satisfy condition (\ref{condneq2}%
) and the quantities $r$, $\bar{\Omega}_{i}$, and $r_{\alpha }$ are
specified in formulas (\ref{3.14}), (\ref{3.15a}), and (\ref{ra}),
respectively. Next, we construct the functional
\begin{align}
\bar{S}^{\prime }=\bar{S}_{T}+\bar{S}_{n}=\int d^{4}x\Big[& -\tfrac{1}{4}%
F_{\mu \nu }F^{\mu \nu }+\tfrac{1}{2}g^{2}k_{AB}n^{A}n^{B}A_{\mu }A^{\mu }-g%
\mathcal{V}(r,\bar{\Omega}_{i},r_{\alpha })  \notag \\
& +\tfrac{1}{2}k_{AB}\big(D_{\mu }\varphi ^{A}-2gn^{A}A_{\mu }\big)D^{\mu
}\varphi ^{B}  \notag \\
& +A_{\mu }^{\ast }\partial ^{\mu }\eta +g\varphi _{A}^{\ast }\big(%
T^{AB}k_{BC}\varphi ^{C}+n^{A}\big)\eta \Big].  \label{comp16}
\end{align}%
We observe that (\ref{comp16}) is nothing but our deformed solution (\ref%
{defsolmast1}) where we implement (\ref{3.19a}). Consequently, equation (\ref%
{Sdef}) ensures that $(\bar{S}^{\prime },\bar{S}^{\prime })=0$. Then, the
pieces of antifield number $0$ and respectively $1$ from (\ref{comp16}) lead
precisely to the Lagrangian action (\ref{3.17}) and gauge transformations (%
\ref{3.18}) obtained in the previous section with the particular choice (\ref%
{3.19a})
\begin{align}
\bar{S}_{0}^{\prime \mathrm{L}}\big[A^{\mu },\varphi ^{A}\big]=\int d^{4}x%
\Big[& -\tfrac{1}{4}F_{\mu \nu }F^{\mu \nu }+\tfrac{1}{2}%
g^{2}k_{AB}n^{A}n^{B}A_{\mu }A^{\mu }-g\mathcal{V}(r,\bar{\Omega}%
_{i},r_{\alpha })  \notag \\
& +\tfrac{1}{2}k_{AB}\big(D_{\mu }\varphi ^{A}-2gn^{A}A_{\mu }\big)D^{\mu
}\varphi ^{B}\Big],  \label{comp17a}
\end{align}%
\begin{equation}
\bar{\delta}_{\epsilon }A^{\mu }=\partial ^{\mu }\epsilon ,\qquad \bar{\delta%
}_{\epsilon }\varphi ^{A}=g\big(T_{\hphantom{A}B}^{A}\varphi ^{B}+n^{A}\big)%
\epsilon ,  \label{comp17b}
\end{equation}%
which indeed emphasize a gauge theory with a massive vector field.

The last arguments enable us to state the following conclusions: (b) our
method in the presence of the starting point (\ref{comp1}) and (\ref{comp3})
is conceptually different from the Higgs mechanism; (c) it is applicable to
an arbitrary gauge-invariant function $V_{1}(\varphi ^{A})$, which is no
longer constrained to display an absolute minimum.

At this stage, we remark that the final outputs of the Abelian Higgs
mechanism ((\ref{higs4}) and (\ref{higs5})) and those of our procedure ((\ref%
{comp17a}) and (\ref{comp17b})) are different in general. In the sequel we
investigate whether our method is capable of rendering the results of the
Abelian Higgs mechanism. In view of this, we take $n^{A}$ of the form
\begin{equation}
n^{A}=T_{\hphantom{A}B}^{A}v_{0}^{B}.  \label{new1}
\end{equation}%
In this situation, equations (\ref{ra}) become
\begin{equation}
\frac{\partial r_{\alpha }}{\partial \varphi ^{A}}T_{\hphantom{A}B}^{A}\big(%
\varphi ^{B}+v_{0}^{B}\big)=0  \label{higs5.2}
\end{equation}%
and, by virtue of (\ref{higs5.1}), obviously admit at least the solution
\begin{equation}
r_{\alpha }\rightarrow r_{1}=V_{1}^{\mathrm{Higgs}}(\varphi ^{A}+v_{0}^{A}),
\label{new2}
\end{equation}%
which allows us to choose $\mathcal{V}(r,\bar{\Omega}_{i},r_{\alpha })$ like
\begin{equation}
\mathcal{V}(r,\bar{\Omega}_{i},r_{\alpha })\rightarrow \mathcal{V}(r_{1})=%
\frac{1}{g}r_{1}=\frac{1}{g}V_{1}^{\mathrm{Higgs}}(\varphi ^{A}+v_{0}^{A}).
\label{new3}
\end{equation}%
Now, we particularize the procedure developed between formulas (\ref{comp8})
and (\ref{comp17b}) to the case where
\begin{equation}
V_{1}(\varphi ^{A})=V_{1}^{\mathrm{Higgs}}(\varphi ^{A})  \label{comp4.1}
\end{equation}%
and $n^{A}$ together with $\mathcal{V}$ are expressed by (\ref{new1}) and (%
\ref{new3}). The ansatz described by formula (\ref{comp4.1}) leads to the
fact that relations (\ref{comp1}) and (\ref{comp3}) precisely reduce to (\ref%
{compX}) and (\ref{compX1}). Therefore, the solution to the master equation
corresponding to the gauge theory described by formulas (\ref{compX}) and (%
\ref{compX1}) is given by
\begin{align}
\bar{S}_{T,\mathrm{Higgs}}=\int d^{4}x\Big[& -\tfrac{1}{4}F_{\mu \nu }F^{\mu
\nu }+\tfrac{1}{2}k_{AB}\big(D_{\mu }\varphi ^{A}\big)D^{\mu }\varphi ^{B}
\notag \\
& -V_{1}^{\mathrm{Higgs}}(\varphi ^{A})+A_{\mu }^{\ast }\partial ^{\mu }\eta
+g\varphi _{A}^{\ast }T_{\hphantom{A}B}^{A}\varphi ^{B}\eta \Big]
\label{Sthigs}
\end{align}%
and satisfies by construction the equation
\begin{equation}
(\bar{S}_{T,\mathrm{Higgs}},\bar{S}_{T,\mathrm{Higgs}})=0.  \label{compX2}
\end{equation}%
The role of the functional (\ref{comp12}) is played here by
\begin{align}
\bar{S}_{v_{0}}=\int d^{4}x\Big(& \tfrac{1}{2}g^{2}k_{AB}T_{\hphantom{A}%
C}^{A}v_{0}^{C}T_{\hphantom{B}D}^{B}v_{0}^{D}A_{\mu }A^{\mu }-gk_{AB}T_{%
\hphantom{A}C}^{A}v_{0}^{C}A_{\mu }D^{\mu }\varphi ^{B}  \notag \\
& +V_{1}^{\mathrm{Higgs}}(\varphi ^{A})-V_{1}^{\mathrm{Higgs}}(\varphi
^{A}+v_{0}^{A})+g\varphi _{A}^{\ast }T_{\hphantom{A}C}^{A}v_{0}^{C}\eta \Big)%
.  \label{compX3}
\end{align}%
Functional (\ref{compX3}) follows from (\ref{comp12}) where we set (\ref%
{new1}), (\ref{new3}), and (\ref{comp4.1}). By direct computation, we infer
that
\begin{equation}
(\bar{S}_{v_{0}},\bar{S}_{v_{0}})\neq 0.  \label{compX4}
\end{equation}%
Finally, we construct the functional (that results from (\ref{comp16}) where
we use choices (\ref{new1}) and (\ref{new3}))
\begin{align}
\bar{S}_{\mathrm{Higgs}}^{\prime }& =\bar{S}_{T,\mathrm{Higgs}}+\bar{S}%
_{v_{0}}  \notag \\
& =\int d^{4}x\Big[-\tfrac{1}{4}F_{\mu \nu }F^{\mu \nu }+\tfrac{1}{2}%
g^{2}k_{AB}T_{\hphantom{A}C}^{A}v_{0}^{C}T_{\hphantom{B}D}^{B}v_{0}^{D}A_{%
\mu }A^{\mu }  \notag \\
& +\tfrac{1}{2}k_{AB}\big(D_{\mu }\varphi ^{A}-2gT_{\hphantom{A}%
C}^{A}v_{0}^{C}A_{\mu }\big)D^{\mu }\varphi ^{B}-V_{1}^{\mathrm{Higgs}%
}(\varphi ^{A}+v_{0}^{A})  \notag \\
& +A_{\mu }^{\ast }\partial ^{\mu }\eta +g\varphi _{A}^{\ast }T_{\hphantom{A}%
B}^{A}\big(\varphi ^{B}+v_{0}^{B}\big)\eta \Big],  \label{compX5}
\end{align}%
which obviously verifies the master equation
\begin{equation}
(\bar{S}_{\mathrm{Higgs}}^{\prime },\bar{S}_{\mathrm{Higgs}}^{\prime })=0.
\label{compX6}
\end{equation}%
The projection of (\ref{compX5}) on antifield number $0$ provides the
Lagrangian action
\begin{align}
& \bar{S}_{0,\mathrm{Higgs}}^{\prime \mathrm{L}}\big[A^{\mu },\varphi ^{A}%
\big]=\int d^{4}x\Big[-\tfrac{1}{4}F_{\mu \nu }F^{\mu \nu }+\tfrac{1}{2}%
g^{2}k_{AB}T_{\hphantom{A}C}^{A}T_{\hphantom{B}D}^{B}v_{0}^{C}v_{0}^{D}A_{%
\mu }A^{\mu }  \notag \\
& +\tfrac{1}{2}k_{AB}\big(D_{\mu }\varphi ^{A}-2gT_{\hphantom{A}%
C}^{A}v_{0}^{C}A_{\mu }\big)D^{\mu }\varphi ^{B}-V_{1}^{\mathrm{Higgs}%
}(\varphi ^{A}+v_{0}^{A})\Big],  \label{new4}
\end{align}%
while from the terms of antifield number $1$ we read the gauge
transformations of (\ref{new4}) like
\begin{equation}
\bar{\delta}_{\epsilon }A^{\mu }=\partial ^{\mu }\epsilon ,\qquad \bar{\delta%
}_{\epsilon }\varphi ^{A}=gT_{\hphantom{A}B}^{A}\big(\varphi ^{B}+v_{0}^{B}%
\big)\epsilon .  \label{new5}
\end{equation}%
It is obvious that (\ref{new4}) and (\ref{new5}) are nothing but (\ref{higs4}%
) and (\ref{higs5}) modulo the identification
\begin{equation}
\varphi ^{A}\longleftrightarrow \tilde{\varphi}^{A}.  \label{new6}
\end{equation}%
Formulas (\ref{new4})--(\ref{new6}) argue that: (d) in the case described by
relations (\ref{new1}), (\ref{new3}), and (\ref{comp4.1}), the results of
our procedure do indeed coincide with those the Abelian Higgs mechanism.

In this way, conclusions (a)--(d) obtained in this section prove that our
procedure may be regarded as a \emph{cohomological extension} of the Abelian
Higgs mechanism, which is precisly result (vi) announced in the introduction.

\section{BRST interpretation of the Higgs mechanism\label{interpret}}

Formulas (\ref{higs4}) together with (\ref{higs5}) and (\ref{new4})
accompanied by (\ref{new5}) respectively --- in the presence of (\ref{new6})
--- emphasize that the final output of the Abelian Higgs mechanism can be
obtained in two different manners: either by performing the shift
transformations (\ref{higs3}) or by means of the procedure exposed between
formulas (\ref{Sthigs}) and (\ref{new5}). Therefore, we can view the Abelian
Higgs mechanism like the passage from formulas (\ref{compX}) and (\ref%
{compX1}) to (\ref{higs4}) and (\ref{higs5}), or, equivalently, from
relations (\ref{compX}) and (\ref{compX1}) to (\ref{new4}) and (\ref{new5}).

Now, we are in the position to give an interpretation of the Abelian Higgs
mechanism in terms of the antifield-BRST symmetry. In view of this, we adopt
the second manner exposed in the above. Actually, the passage from (\ref%
{compX}) and (\ref{compX1}) to (\ref{new4}) and (\ref{new5}) means, at the
level of the BRST formalism, the transit from (\ref{Sthigs})--(\ref{compX2})
to (\ref{compX5})--(\ref{compX6}). On the one hand, formulas (\ref{Sthigs})
and (\ref{compX2}) define a differential of ghost number equal to $1$ that
acts like
\begin{equation}
\bar{s}_{T,\mathrm{Higgs}}F=(F,\bar{S}_{T,\mathrm{Higgs}}),\qquad \bar{s}_{T,%
\mathrm{Higgs}}^{2}=0.  \label{comp10}
\end{equation}%
The operator $\bar{s}_{T,\mathrm{Higgs}}$ signifies the BRST differential
associated with the theory governed by (\ref{compX}) and (\ref{compX1}). On
the other hand, relations (\ref{compX5}) and (\ref{compX6}) define also a
differential of ghost number equal to $1$, via
\begin{equation}
\bar{s}_{\mathrm{Higgs}}^{\prime }F=(F,\bar{S}_{\mathrm{Higgs}}^{\prime
}),\qquad \bar{s}_{\mathrm{Higgs}}^{\prime 2}=0,  \label{comp10a}
\end{equation}%
which is nothing but the BRST differential corresponding to the gauge theory
pictured by (\ref{new4}) and (\ref{new5}). At the same time, formulas (\ref%
{compX3}) and (\ref{compX4}) induce an odd derivation of ghost number $1$
\begin{equation}
\bar{s}_{v_{0}}F=(F,\bar{S}_{v_{0}}),\qquad \bar{s}_{v_{0}}^{2}\neq 0.
\label{comp10b}
\end{equation}%
By means of definitions (\ref{comp10})--(\ref{comp10b}), relation (\ref%
{compX5}) connects these three operators through
\begin{equation}
\bar{s}_{\mathrm{Higgs}}^{\prime }=\bar{s}_{T,\mathrm{Higgs}}+\bar{s}%
_{v_{0}}.  \label{comp10c}
\end{equation}%
Obviously, the operators from (\ref{comp10c}) act on the same BRST algebra,
such that we find that
\begin{equation}
H^{k}(\bar{s}_{\mathrm{Higgs}}^{\prime })\neq H^{k}(\bar{s}_{T,\mathrm{Higgs}%
}),\qquad k\geqslant 0,  \label{comp10d}
\end{equation}%
where $H^{k}(\bar{s}_{\mathrm{Higgs}}^{\prime })$ and $H^{k}(\bar{s}_{T,%
\mathrm{Higgs}})$ represent the cohomologies of $\bar{s}_{\mathrm{Higgs}%
}^{\prime }$ and $\bar{s}_{T,\mathrm{Higgs}}$ in ghost number $k$ computed
in the space of local functionals. In particular, (\ref{comp10d}) leads to
\begin{equation}
H^{0}(\bar{s}_{\mathrm{Higgs}}^{\prime })\neq H^{0}(\bar{s}_{T,\mathrm{Higgs}%
}),  \label{comp10e}
\end{equation}%
which further implies that the classical observables associated with the
theories described by relations (\ref{compX}) and (\ref{compX1}) and
respectively (\ref{new4}) and (\ref{new5}) are different. We recall that the
classical observables of a given gauge theory are gauge-invariant local
functionals modulo the field equations. In conclusion, the passage from (\ref%
{compX}) and (\ref{compX1}) to (\ref{new4}) and (\ref{new5}), which we have
seen that represents a proper description of the Higgs mechanism, means the
transit from the BRST differential $\bar{s}_{T,\mathrm{Higgs}}$ to the BRST
differential $\bar{s}_{\mathrm{Higgs}}^{\prime }$ (using relation (\ref%
{comp10c})), which implies that the classical observables of these two
theories are different. The last statement stays at the core of the
interpretation of the Abelian Higgs mechanism in the light of the BRST
symmetry and meanwhile proves result (vii) from Section \ref{intro}. In this
context we remark that the role of the (scalar) shift transformations (\ref%
{higs3}) from the traditional approach to the Higgs mechanism is played in
the framework of the BRST symmetry by the odd derivation (\ref{comp10b}).
Thus, all the main objectives of this paper have been accomplished.

\section{Examples\label{examples}}

In this section we will exemplify the general results obtained in Section %
\ref{inter} to the case of the interactions between a vector field and one,
two, and three real scalar fields. In view of this, from now on we work with
\begin{equation}
k_{AB}=\delta _{AB},\qquad \mathcal{V}(r,\bar{\Omega}_{i},r_{\alpha
})\rightarrow \mathcal{V}(r,\bar{\Omega}_{i})=c_{1}r+c_{2}\big(r+\tfrac{1}{2}%
b_{i}\bar{\Omega}_{i}^{2}\big)^{2}+\tfrac{1}{2}d_{i}\bar{\Omega}_{i}^{2},
\label{3.22}
\end{equation}%
where $c_{1}$, $c_{2}$, $b_{i}$, and $d_{i}$ represent some arbitrary real
constants. On account of the first choice from (\ref{3.22}) all scalar
indices $A$, $B$, $C$, and so on will be set in lower positions. Although $%
\mathcal{V}(r,\bar{\Omega}_{i})$ can be taken of a more general form, here
we work with a polynomial expression, of the form (\ref{3.22}), in order to
emphasize other interesting aspects of our procedure.

\subsection{The case of one scalar field}

First, we consider the case $N_{0}=1$, which corresponds to the interactions
between a vector field and a single real scalar field, to be denoted by $%
\varphi _{1}\equiv \varphi $. Due to the antisymmetry property of the
arbitrary constants $T_{AB}\rightarrow T_{11}$, the only possible choice is $%
T_{11}=0$. The fact that $\mathrm{rank}(T_{11})=0$ does not contradict
condition (\ref{condneq1}) since here $N_{0}=1$. Obviously, the null vectors
$\tau _{\hphantom{B}i}^{B}$ are absent. Then, by means of (\ref{3.14}) and
of the notation $n_{A}\rightarrow n_{1}\equiv n\neq 0$, formula (\ref{3.22})
leads to the fact that $\mathcal{V}(r,\bar{\Omega}_{i})\rightarrow \mathcal{V%
}(r) =(1/2)n^{2}(c_{1} +\tfrac{1}{2}c_{2}n^{2})$, so it reduces in this
situation to an irrelevant constant and will therefore be omitted.
Consequently, formulas (\ref{3.17}) and (\ref{3.18}) in the presence of
choice (\ref{3.19a}) become
\begin{gather}
\bar{S}_{0}^{\mathrm{L}}[A^{\mu },\varphi ]=\int d^{4}x\Big[-\tfrac{1}{4}%
F_{\mu \nu }F^{\mu \nu }+\tfrac{1}{2}(gnA_{\mu } -\partial _{\mu } \varphi
)(gnA^{\mu }-\partial ^{\mu }\varphi )\Big],  \label{3.23} \\
\bar{\delta}_{\epsilon }A^{\mu }=\partial ^{\mu }\epsilon ,\qquad \bar{\delta%
}_{\epsilon }\varphi =gn\epsilon ,  \label{3.24}
\end{gather}
whereas the gauge-fixed action (\ref{3.21}) with the same choice takes the
particular form
\begin{align}
\bar{S}_{K}=\int d^{4}x\Big[& -\tfrac{1}{4}F_{\mu \nu }F^{\mu \nu }+\tfrac{1%
}{2}g^{2}n^{2}A_{\mu }A^{\mu }-\tfrac{1}{2\xi }(\partial _{\mu } A^{\mu
})^{2}  \notag \\
& +\tfrac{1}{2}(\partial _{\mu }\varphi )\partial ^{\mu }\varphi -\tfrac{1}{2%
}\xi g^{2}n^{2}\varphi ^{2} +(\partial _{\mu }\bar{\eta}) \partial ^{\mu }
\eta -\xi g^{2}n^{2}\bar{\eta}\eta \Big].  \label{3.25}
\end{align}
Analyzing relations (\ref{3.23}) and (\ref{3.24}), we observe that they
provide nothing but the Stueckelberg coupling between a vector field and a
scalar field $\varphi $. We emphasized in the general context from Section %
\ref{inter} that the unphysical scalar degree of freedom is $n_{A}\varphi
_{A}\rightarrow n\varphi $, so the only scalar field from the present
context, $\varphi $, describes no physical degree of freedom. Therefore, the
gauge-fixed action (\ref{3.25}) comprises three physical degrees of freedom
associated with the massive vector field $A^{\mu }$, as well as the
unphysical degrees of freedom corresponding to $\{\varphi ,\bar{\eta},\eta
\} $. In this particular situation the ghosts are not coupled to the
Stueckelberg scalar (since $T_{11}=0$).

\subsection{The case of two scalar fields}

Second, we analyze the case $N_{0}=2$, i.e., the interactions among a vector
field and two real scalar fields. We take the elements $T_{AB}$ and the
constants $n_{A}$ of the form
\begin{equation}
T_{AB}=\beta \left(
\begin{array}{cc}
0 & 1 \\
-1 & 0%
\end{array}
\right) ,\qquad n_{1}=0,\qquad n_{2}\equiv -n,  \label{3.26}
\end{equation}
with both $\beta $ and $n$ nonvanishing. It is easy to see that in this case
there are also no nontrivial null vectors $\tau _{\hphantom{B}i}^{B}$, so
the dependence of $\mathcal{V}$ on $\bar{\Omega}_{i}$ is trivial. Then, (\ref%
{3.22}) reduces to $\mathcal{V}(r)=c_{1}r+c_{2}r^{2}$. As a consequence,
from expressions (\ref{3.17}) and (\ref{3.18}) where we set (\ref{3.19a}) we
generate the interacting Lagrangian action and accompanying gauge
transformations in this case like
\begin{align}
\bar{S}_{0}^{\mathrm{L}}[A^{\mu },\varphi _{1},\varphi _{2}]=\int d^{4}x %
\Big\{& -\tfrac{1}{4}F_{\mu \nu }F^{\mu \nu } +\tfrac{1}{2}g^{2}n^{2}A_{\mu
}A^{\mu }  \notag \\
& +\tfrac{1}{2}(\partial _{\mu }\varphi _{1})\partial ^{\mu }\varphi _{1} -%
\tfrac{1}{2}g\beta ^{2}\big(c_{1}+3n^{2}c_{2}\big)\varphi _{1}^{2}  \notag \\
& +\tfrac{1}{2}(\partial _{\mu }\varphi _{2})\partial ^{\mu }\varphi _{2} -%
\tfrac{1}{2}g\beta ^{2}\big(c_{1}+n^{2}c_{2}\big)\varphi _{2}^{2}  \notag \\
& -\tfrac{1}{4}gc_{2}\beta ^{3}\big(\varphi _{1}^{2}+\varphi _{2}^{2}\big) %
\big[\beta \big(\varphi _{1}^{2}+\varphi _{2}^{2}\big)+4n\varphi _{1}\big]
\notag \\
& +g\beta A_{\mu }\big(\varphi _{1}\partial ^{\mu }\varphi _{2}-\varphi
_{2}\partial ^{\mu }\varphi _{1}\big)+gnA_{\mu }\partial ^{\mu }\varphi _{2}
\notag \\
& -g\beta n\big(c_{1}+n^{2}c_{2}\big)\varphi _{1}  \notag \\
& +\tfrac{1}{2}g^{2}\beta \big[\beta \big(\varphi _{1}^{2}+\varphi _{2}^{2}%
\big)+2n\varphi _{1}\big]A_{\mu }A^{\mu }\Big\},  \label{3.27}
\end{align}
\begin{equation}
\bar{\delta}_{\epsilon }A^{\mu }=\partial ^{\mu }\epsilon ,\qquad \bar{\delta%
}_{\epsilon }\varphi _{1}=g\beta \varphi _{2}\epsilon ,\qquad \bar{\delta}%
_{\epsilon }\varphi _{2}=-g(\beta \varphi _{1}+n)\epsilon .  \label{3.28}
\end{equation}
We remark that the real constants $c_{1}$, $c_{2}$, $\beta $, and $n$
appearing in (\ref{3.27}) and (\ref{3.28}) are arbitrary (with $\beta $ and $%
n$ nonvanishing), such that the mass of the vector field does not depend
either on $c_{1}$, $c_{2}$, or $\beta $. Relations (\ref{3.27}) and (\ref%
{3.28}) represent the most general expressions (taking (\ref{3.19a}) into
consideration) that describe an interacting gauge theory with one massive
vector field and two scalars. We remark that formulas (\ref{7H}) and (\ref%
{8H}) follow from (\ref{3.27}) and (\ref{3.28}) in the particular case
\begin{equation}
g=q,\qquad gc_{1}=\mu ^{2},\qquad gc_{2}=\tfrac{1}{4}\lambda >0,\qquad \beta
=1,\qquad n=v.  \label{parthiggs1}
\end{equation}
Obviously, the results emerging from the Abelian Higgs mechanism are
obtained from (\ref{3.27})--(\ref{parthiggs1}) in the more particular
situation
\begin{equation}
\mu ^{2}<0,\qquad v=\sqrt{\frac{-4\mu ^{2}}{{\lambda }}}.  \label{parthiggs2}
\end{equation}

The gauge-fixed action (\ref{3.21}) where we put (\ref{3.19a}) takes (for
this example) the concrete expression
\begin{align}
\bar{S}_{K}= \int d^{4}x \Big\{& -\tfrac{1}{4} F_{\mu \nu } F^{\mu \nu } +%
\tfrac{1}{2} g^{2} n^{2} A_{\mu } A^{\mu } -\tfrac{1}{2\xi } (\partial _{\mu
}A^{\mu })^{2}  \notag \\
& +\tfrac{1}{2} (\partial _{\mu } \varphi _{1}) \partial ^{\mu } \varphi
_{1} -\tfrac{1}{2} g \beta ^{2} \big( c_{1} +3n^{2} c_{2} \big) \varphi
_{1}^{2}  \notag \\
& +\tfrac{1}{2} (\partial _{\mu } \varphi _{2}) \partial ^{\mu }\varphi _{2}
-\tfrac{1}{2} g \big[ \beta ^{2} c_{1} +n^{2} (\beta ^{2}c_{2} +\xi g) \big] %
\varphi _{2}^{2}  \notag \\
& -\tfrac{1}{4} g c_{2} \beta ^{3} \big( \varphi _{1}^{2} +\varphi _{2}^{2} %
\big) \big[ \beta \big( \varphi _{1}^{2} +\varphi _{2}^{2} \big) +4n \varphi
_{1}\big]  \notag \\
& +g \beta A_{\mu } \big( \varphi _{1} \partial ^{\mu } \varphi _{2}-
\varphi _{2} \partial ^{\mu } \varphi _{1} \big) -g \beta n \big( c_{1}
+n^{2} c_{2} \big) \varphi _{1}  \notag \\
& +\tfrac{1}{2} g^{2} \beta \big[ \beta \big( \varphi _{1}^{2} +\varphi
_{2}^{2} \big) +2n \varphi _{1} \big] A_{\mu } A^{\mu } +(\partial _{\mu }
\bar{\eta}) \partial ^{\mu } \eta  \notag \\
& -\xi g^{2} n^{2} \bar{\eta} \eta -\xi g^{2} \beta n \varphi _{1} \bar{\eta}
\eta \Big\}.  \label{3.29}
\end{align}
Here, the unphysical scalar degree of freedom is $n_{A}\varphi
_{A}\rightarrow -n\varphi _{2}$ and hence it reduces precisely to the scalar
field $\varphi _{2}$. Accordingly, this model exhibits four physical degrees
of freedom (three corresponding to the massive vector field $A^{\mu }$ and
one associated with $\varphi _{1}$) and the unphysical degrees of freedom $%
\{ \varphi _{2},\bar{\eta},\eta \} $. Moreover, the vertex $(-)\xi
g^{2}\beta n\varphi _{1}\bar{\eta}\eta $ from (\ref{3.29}) signalizes that
the physical scalar $\varphi _{1}$ is coupled to the ghosts. This vertex,
omitted in QFT textbooks, should be present also in the gauge-fixed action
resulting from the Abelian Higgs mechanism, which follows from (\ref{3.29})
with the choices (\ref{parthiggs1}) and (\ref{parthiggs2}). Its presence is
important since it ensures the invariance of the gauge-fixed action (\ref%
{3.29}) under the gauge-fixed BRST transformations (\ref{3.21a}) and (\ref%
{3.21b}) particularized to this example, which is otherwise lost.

\subsection{The case of three scalar fields}

Third, we investigate the case $N_{0}=3$, which provides the interactions
among a vector field and three real scalar fields. In this situation we take
\begin{equation}
T_{AB}=\beta \left(
\begin{array}{ccc}
0 & 1 & 0 \\
-1 & 0 & 0 \\
0 & 0 & 0%
\end{array}
\right) ,  \label{3.30}
\end{equation}
with $\beta $ nonvanishing. We remark that matrix (\ref{3.30}) possesses the
null vector
\begin{equation}
\tau _{1}=0,\qquad \tau _{2}=0,\qquad \tau _{3}=1.  \label{null}
\end{equation}

First, we choose $n_{A}$'s of the form
\begin{equation}
n_{1}=0,\qquad n_{2}=0,\qquad n_{3}\equiv n,  \label{n1}
\end{equation}%
with $n$ nonvanishing. With the help of (\ref{null}) and (\ref{n1}) we
observe that (\ref{iffnon}) is not satisfied, so the dependence of $\mathcal{%
V}$ on $\bar{\Omega}_{i}$ is again trivial, such that (\ref{3.22}) reduces
to $\mathcal{V}(r)=c_{1}r+c_{2}r^{2}$. In this context formulas (\ref{3.17})
and (\ref{3.18}) take the particular form (being understood that we
implement (\ref{3.19a}))
\begin{align}
\bar{S}_{0}^{\mathrm{L}}[A^{\mu },\varphi _{1},\varphi _{2},\varphi
_{3}]=\int d^{4}x\Big[& -\tfrac{1}{4}F_{\mu \nu }F^{\mu \nu } +\tfrac{1}{2}%
g^{2}n^{2}A_{\mu }A^{\mu }  \notag \\
& +\tfrac{1}{2}(\partial _{\mu }\varphi _{1})\partial ^{\mu }\varphi _{1} +%
\tfrac{1}{2}(\partial _{\mu }\varphi _{2})\partial ^{\mu }\varphi _{2}
\notag \\
& +\tfrac{1}{2}(\partial _{\mu }\varphi _{3})\partial ^{\mu }\varphi _{3} -%
\tfrac{1}{2}g\beta ^{2}\big(c_{1}+n^{2}c_{2}\big)\big(\varphi
_{1}^{2}+\varphi _{2}^{2}\big)  \notag \\
& -\tfrac{1}{4}gc_{2}\beta ^{4}\big(\varphi _{1}^{2}+\varphi _{2}^{2}\big) %
^{2}+g\beta A_{\mu }\big(\varphi _{1}\partial ^{\mu }\varphi _{2}-\varphi
_{2}\partial ^{\mu }\varphi _{1}\big)  \notag \\
& -gnA_{\mu }\partial ^{\mu }\varphi _{3}+\tfrac{1}{2}g^{2}\beta ^{2}\big( %
\varphi _{1}^{2}+\varphi _{2}^{2}\big)A_{\mu }A^{\mu }\Big],  \label{3.31}
\end{align}
\begin{equation}
\bar{\delta}_{\epsilon }A^{\mu }=\partial ^{\mu }\epsilon ,\qquad \bar{\delta%
}_{\epsilon }\varphi _{1}=g\beta \varphi _{2}\epsilon ,\qquad \bar{\delta}%
_{\epsilon }\varphi _{2}=-g\beta \varphi _{1}\epsilon ,\qquad \bar{\delta}%
_{\epsilon }\varphi _{3}=gn\epsilon .  \label{3.32}
\end{equation}
Again, the mass of the vector field does not depend either on $c_{1}$, $%
c_{2} $, or $\beta $. If we choose the constants $c_{1}$ and $c_{2}$ such
that $g\big(c_{1}+n^{2}c_{2}\big)>0$ and set $\beta =1$, then action (\ref%
{3.31}) describes precisely an $U(1)$-type coupling between the massive
physical scalars $\{\varphi _{1},\varphi _{2}\}$ and a massive vector field
in the presence of the (unphysical) Stueckelberg scalar field $\varphi _{3}$%
. Meanwhile, (\ref{3.31}) contains also interactions involving the two
physical scalars. The gauge-fixed action (\ref{3.21}) corresponding to (\ref%
{3.31}) is given by
\begin{align}
\bar{S}_{K}=\int d^{4}x\Big[& -\tfrac{1}{4}F_{\mu \nu }F^{\mu \nu }+\tfrac{1%
}{2}g^{2}n^{2}A_{\mu }A^{\mu }-\tfrac{1}{2\xi }(\partial _{\mu } A^{\mu
})^{2}  \notag \\
& +\tfrac{1}{2}(\partial _{\mu }\varphi _{1})\partial ^{\mu }\varphi _{1}+%
\tfrac{1}{2}(\partial _{\mu }\varphi _{2})\partial ^{\mu }\varphi _{2}+%
\tfrac{1}{2}(\partial _{\mu }\varphi _{3})\partial ^{\mu }\varphi _{3}
\notag \\
& -\tfrac{1}{2}g\beta ^{2}\big(c_{1}+n^{2}c_{2}\big)\big(\varphi
_{1}^{2}+\varphi _{2}^{2}\big)-\tfrac{1}{4}gc_{2}\beta ^{4}\big(\varphi
_{1}^{2}+\varphi _{2}^{2}\big)^{2}  \notag \\
& +g\beta A_{\mu }\big(\varphi _{1}\partial ^{\mu }\varphi _{2}-\varphi
_{2}\partial ^{\mu }\varphi _{1}\big)+\tfrac{1}{2}g^{2}\beta ^{2} \big(%
\varphi _{1}^{2}+\varphi _{2}^{2}\big)A_{\mu }A^{\mu }  \notag \\
& -\tfrac{1}{2}\xi g^{2}n^{2}\varphi _{3}^{2}+(\partial _{\mu }\bar{\eta}%
)\partial ^{\mu }\eta -\xi g^{2}n^{2}\bar{\eta}\eta \Big].  \label{3.33}
\end{align}
This example underlies five physical degrees of freedom (three corresponding
to $A^{\mu }$ and one for each of the scalars $\varphi _{1}$ and
respectively $\varphi _{2}$), whereas $\{\varphi _{3},\bar{\eta},\eta \}$
are unphysical. We notice that the scalar fields are no longer coupled to
the ghosts, like in the first example.

Second, we take $n_{A}$'s of the form
\begin{equation}
n_{1}=0,\qquad n_{2}\equiv -n,\qquad n_{3}=0,  \label{n2}
\end{equation}
with $n$ nonvanishing. By means of (\ref{null}) and (\ref{n2}) we obtain
that (\ref{iffnon}) is satisfied, so $\mathcal{V}$ depends nontrivially on $%
r $ and $\bar{\Omega}_{1}\equiv \varphi _{3}$. Denoting the corresponding
constants $b_{1}$ and $d_{1}$ from (\ref{3.22}) with $b$ and respectively $d$%
, relations (\ref{3.17}) and (\ref{3.18}) (in the presence of (\ref{3.19a}))
become
\begin{align}
\bar{S}_{0}^{\mathrm{L}}[A^{\mu },\varphi _{1},\varphi _{2},\varphi
_{3}]=\int d^{4}x\Big\{& -\tfrac{1}{4}F_{\mu \nu }F^{\mu \nu } +\tfrac{1}{2}%
g^{2}n^{2}A_{\mu }A^{\mu }  \notag \\
& +\tfrac{1}{2}(\partial _{\mu }\varphi _{1})\partial ^{\mu }\varphi _{1} -%
\tfrac{1}{2}g\beta ^{2}\big(c_{1}+3n^{2}c_{2}\big)\varphi _{1}^{2}  \notag \\
& +\tfrac{1}{2}(\partial _{\mu }\varphi _{2})\partial ^{\mu }\varphi _{2} -%
\tfrac{1}{2}g\beta ^{2}\big(c_{1}+n^{2}c_{2}\big)\varphi _{2}^{2}  \notag \\
& +\tfrac{1}{2}(\partial _{\mu }\varphi _{3})\partial ^{\mu }\varphi _{3} -%
\tfrac{1}{2}g\big(d+n^{2}bc_{2}\big)\varphi _{3}^{2}  \notag \\
& -\tfrac{1}{4}gc_{2}\big[\beta ^{2}\big(\varphi _{1}^{2}+\varphi _{2}^{2}%
\big)+b\varphi _{3}^{2}\big]\times  \notag \\
& \times \big[\beta ^{2}\big(\varphi _{1}^{2}+\varphi _{2}^{2}\big)+b\varphi
_{3}^{2}+4\beta n\varphi _{1}\big]  \notag \\
& +g\beta A_{\mu }\big(\varphi _{1}\partial ^{\mu }\varphi _{2}-\varphi
_{2}\partial ^{\mu }\varphi _{1}\big)+gnA_{\mu }\partial ^{\mu }\varphi _{2}
\notag \\
& -g\beta n\big(c_{1}+n^{2}c_{2}\big)\varphi _{1}  \notag \\
& +\tfrac{1}{2}g^{2}\beta \big[\beta \big(\varphi _{1}^{2}+\varphi _{2}^{2}%
\big)+2n\varphi _{1}\big]A_{\mu }A^{\mu }\Big\},  \label{3.34}
\end{align}
\begin{equation}
\bar{\delta}_{\epsilon }A^{\mu }=\partial ^{\mu }\epsilon ,\quad \bar{\delta}%
_{\epsilon }\varphi _{1}=g\beta \varphi _{2}\epsilon ,\quad \bar{\delta}%
_{\epsilon }\varphi _{2}=-g(\beta \varphi _{1}+n)\epsilon ,\quad \bar{\delta}%
_{\epsilon }\varphi _{3}=0.  \label{3.35}
\end{equation}
Like in the previous cases, the mass of the vector field does not depend on
the arbitrary constants $c_{1}$, $c_{2}$, $\beta $, $b$, and $d$. In this
situation the unphysical scalar degree of freedom is $\varphi _{2}$. Among
the five physical degrees of freedom of this model, three correspond to the
massive vector field and two to the scalar fields $\varphi _{1}$ and $%
\varphi _{3}$ (the last is gauge-invariant). It is interesting to notice
that if we take the constants $c_{1}$, $c_{2}$, $b$, and $d$ such that $g%
\big(c_{1}+3n^{2}c_{2}\big)>0$ and $g\big(d+n^{2}bc_{2}\big)>0$, then the
two physical scalars possess in general different masses. At the same time,
in (\ref{3.34}) there are present interactions involving all the scalar
fields. Although gauge-invariant, the scalar field $\varphi _{3}$ is not
coupled to the vector field. The gauge-fixed action (\ref{3.21}) associated
with (\ref{3.34}) takes in the second case the form
\begin{align}
\bar{S}_{K}=\int d^{4}x\Big\{& -\tfrac{1}{4}F_{\mu \nu }F^{\mu \nu }+\tfrac{1%
}{2}g^{2}n^{2}A_{\mu }A^{\mu }-\tfrac{1}{2\xi }(\partial _{\mu } A^{\mu
})^{2}  \notag \\
& +\tfrac{1}{2}(\partial _{\mu }\varphi _{1})\partial ^{\mu }\varphi _{1} -%
\tfrac{1}{2}g\beta ^{2}\big(c_{1}+3n^{2}c_{2}\big)\varphi _{1}^{2}  \notag \\
& +\tfrac{1}{2}(\partial _{\mu }\varphi _{2})\partial ^{\mu }\varphi _{2} -%
\tfrac{1}{2}g\big[\beta ^{2}c_{1}+n^{2}(\beta ^{2}c_{2}+\xi g) \big] \varphi
_{2}^{2}  \notag \\
& +\tfrac{1}{2}(\partial _{\mu }\varphi _{3})\partial ^{\mu }\varphi _{3} -%
\tfrac{1}{2}g\big(d+n^{2}bc_{2}\big)\varphi _{3}^{2}  \notag \\
& -\tfrac{1}{4}gc_{2}\big[\beta ^{2}\big(\varphi _{1}^{2}+\varphi _{2}^{2} %
\big)+b\varphi _{3}^{2}\big]\times  \notag \\
& \times \big[\beta ^{2}\big(\varphi _{1}^{2}+\varphi _{2}^{2}\big)+b\varphi
_{3}^{2}+4\beta n\varphi _{1}\big]  \notag \\
& +g\beta A_{\mu }\big(\varphi _{1}\partial ^{\mu }\varphi _{2}-\varphi
_{2}\partial ^{\mu }\varphi _{1}\big)-g\beta n \big( c_{1}+n^{2}c_{2} \big) %
\varphi _{1}  \notag \\
& +\tfrac{1}{2}g^{2}\beta \big[\beta \big(\varphi _{1}^{2}+\varphi _{2}^{2}%
\big)+2n\varphi _{1}\big]A_{\mu }A^{\mu }  \notag \\
& +(\partial _{\mu }\bar{\eta})\partial ^{\mu }\eta -\xi g^{2}n^{2}\bar{\eta}%
\eta -\xi g^{2}\beta n\varphi _{1}\bar{\eta}\eta \Big\}.  \label{3.36}
\end{align}
At the level of the gauge-fixed action the fields $\{\varphi _{2},\bar{\eta}%
,\eta \}$ obviously describe unphysical degrees of freedom. Like in the
second example, in (\ref{3.36}) there appears a vertex that couples the
physical scalar $\varphi _{1}$ to the ghosts.

\section{Conclusion\label{concl}}

To conclude with, in this paper we developed a novel mass generation
mechanism for an Abelian vector field. This mechanism is based on the
construction of a class of gauge theories whose free limit describes one
massless vector field and a set of massless real scalar fields by means of
the antifield-BRST deformation method. In this setting it was proved that:

\begin{enumerate}
\item The vector field gains mass irrespective of the number of scalar
fields from the collection;

\item The gauge transformations are deformed with respect to the free limit,
but their gauge algebra remains Abelian;

\item The massive vector field propagator behaves like that from the
massless case in the limit of large Euclidean momenta;

\item Our procedure represents a cohomological extension of the Higgs
mechanism;

\item Our scheme reveals an appropriate interpretation of the Higgs
mechanism in the framework of the BRST symmetry.
\end{enumerate}

Our main results were exemplified to the cases where the number of scalar
fields is equal to one, two, and three. The particular situation of
interactions in the presence of two scalar fields strengthens that our
results include those emerging from the Higgs mechanism and, in addition,
lead to a vertex that is omitted in the literature. The same kind of vertex
has also been shown to appear for three scalar fields. The procedure exposed
in this paper opens the perspective towards its generalization to the case
of interactions among a collection of vector fields and a set of real scalar
fields. This problem is under consideration \cite{inprep}.

\end{document}